\documentclass[pre,twocolumn,nobalancelastpage,amsmath,amssymb,showpacs,
nofootinbib,superscriptaddress]{revtex4-1}

\usepackage{graphics}      
\usepackage{graphicx}      
\usepackage{longtable}     
\usepackage{url}           
\usepackage{bm}            
\usepackage{xcolor}
\usepackage{amsmath}
\usepackage{dcolumn}
\usepackage[normalem]{ulem}

\begin{document}
\title{Effect of Collisional Elasticity on the Bagnold Rheology of Sheared Frictionless Two-Dimensional Disks}
\author{Daniel V{\aa}gberg}
\affiliation{Laboratoire Charles Coulomb, UMR 5221 CNRS, Universit{\'e} Montpellier, Montpellier, France
}
\author{Peter Olsson}
\affiliation{Department of Physics, Ume{\aa} University, 901 87 Ume{\aa}, Sweden}
\author{S. Teitel}
\affiliation{Department of Physics and Astronomy, University of Rochester, Rochester, NY 14627}
\date{\today}

\begin{abstract}
We carry out constant volume simulations of steady-state, shear driven flow in a simple model of athermal, bidisperse, soft-core, frictionless disks in two dimensions, using a dissipation law that gives rise to Bagnoldian rheology.  Focusing on the small strain rate limit, we map out the rheological behavior as a function of particle packing fraction $\phi$ and a parameter $Q$ that measures the elasticity of binary particle collisions.  We find a  $Q^*(\phi)$ that marks the clear crossover from a region characteristic of strongly inelastic collisions, $Q<Q^*$, to a region characteristic of weakly inelastic collisions, $Q>Q^*$, and give evidence that $Q^*(\phi)$ diverges as $\phi\to\phi_J$, the shear driven jamming transition.  We thus conclude that the jamming transition at any value of $Q$ behaves the same as the strongly inelastic case, provided one is sufficiently close to $\phi_J$.  We further characterize the differing nature of collisions in the strongly inelastic vs weakly inelastic regions, and recast our results into the constitutive equation form commonly used in discussions of hard granular matter.

\end{abstract}
\maketitle

\section{Introduction}
\label{secIntro}

In a system of athermal ($T=0$) granular particles with soft- or hard-core contact interactions, as the particle packing fraction $\phi$ increases, the system will undergo a jamming transition from a liquid-like flowing state to a rigid but disordered solid state, at a critical packing fraction $\phi_J$ \cite{OHern,Liu+Nagel}.  For particles without intergranular friction, this jamming transition is in general continuous and the transport coefficients characterizing the liquid state response to shear will, in the low strain rate limit $\dot\gamma\to 0$,  diverge continuously as $\phi_J$ is approached from below \cite{OlssonTeitel2007, OlssonTeitelScaling,VOTBagnold1}.  

For the case where the particle contact interaction is the ``spring-dashpot" model \cite{Schafer}, where energy dissipation  is due only to particle collisions and is proportional to the difference in the components of the colliding particles' velocities normal to the surface at the point of contact, 
the system is known \cite{Lois,OH1,OH2,Hatano1,Hatano2,Hatano4,OH3,OlssonTeitelRheology} to display Bagnoldian rheology \cite{Bagnold} with pressure $p$, and shear stress $\sigma$, scaling with shear strain rate $\dot\gamma$ as $p,\sigma\propto\dot\gamma^2$, at sufficiently slow strain rates.
In a recent work \cite{VOTBagnold1} we considered the shear driven jamming transition for such a model of athermal, bidisperse, frictionless, soft-core disks in two dimensions.  We applied a critical scaling analysis to determine the divergence of the Bagnold transport coefficients, $p/\dot\gamma^2$ and $\sigma/\dot\gamma^2$, as one asymptotically approaches $\phi_J$ from below.  Our analysis was for the specific case of a system with strongly inelastic collisions.

In this work we systematically explore the effects on the system rheology when one varies the degree of collision elasticity away from the strongly inelastic limit.  This question was previously considered by Otsuki, Hayakawa and Luding (OHL) \cite{OHL} who argued, from looking at simulations for several specific cases, that the critical scaling associated with the limit of strongly inelastic collisions always exists in a window of $\phi$ sufficiently close to the jamming $\phi_J$, but that the width of this window decreases  as the collisions become increasingly elastic, and ultimately vanishes in the limit of purely elastic collisions.  They used this result to reconcile the behavior of transport coefficients observed in simulations of particles with strongly inelastic collisions, with earlier work on elastically (and nearly-elastically) colliding particles \cite{Luding,Rojo,Khain,Suzuki}. However they did not explicitly determine the location of this crossover from strongly inelastic to nearly elastic behavior, but only presented a schematic picture (their Fig.~18).

In the present work we reexamine this question by exploring rheological behavior over a wide range of packing fraction $\phi$, and a parameter $Q$ that controls the degree of elasticity of particle collisions.  We focus our attention on the hard-core limit of our soft-core particle model, which is attained when the applied shear strain rate $\dot\gamma$ is sufficiently small and so particle overlaps become negligible.  We find that as $Q$ increases at fixed $\phi$, there is a sharp, but non-singular, crossover: at small $Q$ there is a region of behavior characteristic of strongly inelastic collisions, in which transport coefficients are roughly independent of $Q$; at large $Q$ there is a region of behavior characteristic of weakly inelastic collisions, where transport coefficients increase with increasing $Q$ (see Fig.~\ref{BpeBse-vs-Q}).  We explicitly locate this crossover $Q^*(\phi)$ and provide evidence that it diverges as $\phi\to\phi_J$.  Thus a system at any fixed $Q$ is always in the strongly inelastic region $Q<Q^*$ if  one is sufficiently close to $\phi_J$.  This result thus supports the conclusions of OHL  \cite{OHL}.  

The remainder of our paper is organized as follows.  In Sec.~\ref{s2} we present our numerical model and dimensionless variables, describe the calculation of the different pieces of the pressure tensor and corresponding Bagnold transport coefficients, and give details of our numerical simulation method.  In Sec.~\ref{s3} we present our numerical results for the Bagnold coefficients, determine the crossover $Q^*(\phi)$, discuss the implications for the jamming transition as a function of $Q$, and discuss the effect of varying $Q$ on the macroscopic friction $\mu=\sigma/p$.  We also discuss the different behavior of the strongly inelastic vs the weakly inelastic region with regard to the impact angle and time scales of collisions, as well as the average particle contact number $\langle Z\rangle$.  Finally we recast our results into the form of the ``constitutive equations" commonly used to discuss shear flow in systems of hard-core granular  particles \cite{daCruz, Pouliquen,Forterre,Lemaitre, Roux}.
In Sec.~\ref{s4} we summarize our conclusions.

\section{Model and Simulation Method}
\label{s2}

\subsection{Model}

We use a well studied model \cite{OHern} of frictionless, bidisperse, soft-core circular disks in two dimensions, with equal numbers of big and small particles with diameter ratio $d_b/d_s=1.4$.  Particles interact only when they come into contact,  in which case they repel with an elastic potential,
\begin{equation}
{\cal V}_{ij}(r_{ij})=\left\{
\begin{array}{cc}
\frac{1}{\alpha} k_e\left(1-r_{ij}/d_{ij}\right)^\alpha,&r_{ij}<d_{ij}\\
0,&r_{ij}\ge d_{ij}.
\end{array}
\right.
\label{eInteraction}
\end{equation}
Here $r_{ij}\equiv |\mathbf{r}_{ij}|$, where $\mathbf{r}_{ij}\equiv \mathbf{r}_i-\mathbf{r}_j$ is the center to center displacement from particle $j$ at position $\mathbf{r}_j$ to particle $i$ at $\mathbf{r}_i$, and $d_{ij}\equiv (d_i+d_j)/2$ is the average of their diameters.  In this work we will use the value $\alpha=2$, corresponding to a harmonic repulsion.
The resulting elastic force on particle $i$ from particle $j$ is,
\begin{equation}
\mathbf{f}_{ij}^\mathrm{el}=-\dfrac{d{\cal V}_{ij}(r_{ij})}{d\mathbf{r}_i} =\frac{k_e}{d_{ij}} \left(1-\frac{r_{ij}}{d_{ij}}\right)^{\alpha-1} \mathbf{\hat r}_{ij},
\label{efel}
\end{equation}
where $\mathbf{\hat r}_{ij}\equiv \mathbf{r}_{ij}/r_{ij}$ is the inward pointing normal direction at the surface of particle $i$.

Particles also experience a dissipative force when they come into contact.  We take this force to be proportional to the projection of the velocity difference of the contacting particles onto the direction normal to the surface at the point of contact.  The dissipative force on particle $i$ from particle $j$ is, 
\begin{equation}
\mathbf{f}_{ij}^\mathrm{dis}=-k_d [(\mathbf{v}_i-\mathbf{v}_j)\cdot\mathbf{\hat r}_{ij}]\mathbf{\hat r}_{ij},
\label{efdis}
\end{equation}
where $\mathbf{v}_i\equiv d\mathbf{r}_i/dt$ is the center of mass velocity of particle $i$.  We have earlier \cite{OlssonTeitelRheology} denoted this model of dissipation as CD$_{n}$ for ``normal contact dissipation."  This dissipative force is well known to result in Bagnoldian rheology \cite{Lois,OH1,OH2,Hatano1,Hatano2,Hatano4,OH3,OlssonTeitelRheology,daCruz}.  The combination of elastic and dissipative forces of Eqs.~(\ref{efel}) and (\ref{efdis}) is often referred to as the ``spring-dashpot" model \cite{Schafer}.  We note that the constants $k_e$ and $k_d$, which define the strengths of our forces, have different physical units.

Particle motion is governed by the deterministic Newton's equation,
\begin{equation}
m_i\dfrac{d^2\mathbf{r}_i}{dt^2}=\sum_j\left[\mathbf{f}_{ij}^\mathrm{el}+\mathbf{f}_{ij}^\mathrm{dis}\right],
\label{eEqMotion}
\end{equation}
where $m_i$ is the mass of particle $i$ and the sum is over all particles $j$ in contact with particle $i$.  In this work we take particles to have a mass proportional to their area, i.e., small particles have mass $m_s=\rho_0\pi (d_s/2)^2$ and big particles have mass $m_b=\rho_0\pi (d_b/2)^2$, with $\rho_0$ the mass per area.  We define $m_0=\frac{1}{2}\rho_0 d_s^2$ as a unit of mass \cite{noteMass}. 

The above microscopic dynamics possess two important time scales \cite{OlssonTeitelRheology}, the elastic and dissipative relaxation times,
\begin{equation}
\tau_e\equiv\sqrt{m_0d_s^2/k_e},\qquad \tau_d\equiv m_0/k_d.
\label{etaus}
\end{equation}
The parameter
\begin{equation}
Q\equiv \tau_d/\tau_e=\sqrt{m_0k_e/(k_dd_s)^2}
\end{equation}
measures the degree of elasticity of the collisions. 
For the harmonic interaction  that we use, if we regarded the elastic potential of Eq.~(\ref{eInteraction}) as a spring which did {\em not} break when particles lose contact, then $2\pi\tau_e$ would give the undamped natural period of oscillation, $2\tau_d$ would be the decay time, and $Q$ would be the quality factor.

$Q$ may also be
related to the coefficient of restitution $e$ of a collision.  
For the isolated head-on collision of two particles $i$ and $j$, we have,
\begin{equation}
e=\mathrm{exp}\left[-\pi\middle/\sqrt{4\left(\dfrac{m_{ij}}{m_0}\right)\left(\dfrac{d_s}{d_{ij}}\right)^2 Q^2-1}\,\,\right],
\end{equation}
where $m_{ij}= m_im_j/(m_i+m_j)$ is the reduced mass of the two particles \cite{Schafer}.  When $Q<(d_{ij}/2d_s)\sqrt{m_0/m_{ij}}$, so that the argument of the square root would be negative, the collision is completely inelastic with $e=0$.  For two small particles this happens when $Q<Q_d=0.564$.    Note, however, that in our two dimensional geometry, a collision that is not strictly head-on will result in particles separating after the collision even if $e=0$, since tangential relative motion is not dissipated by the force $\mathbf{f}^\mathrm{dis}_{ij}$ of Eq.~(\ref{efdis}).

Our system consists of a fixed total number particles $N$ in a square box of fixed length $L$.  $L$ is chosen to set the particle packing fraction $\phi$,
\begin{equation}
\phi=\dfrac{\pi N}{2L^2}\left[\left(\dfrac{d_s}{2}\right)^2+\left(\dfrac{d_b}{2}\right)^2\right].
\end{equation}
To apply a uniform shear strain rate $\dot\gamma$ in the $\mathbf{\hat x}$ direction, we use periodic Lees-Edwards boundary conditions \cite{LeesEdwards}, so that a particle at position $\mathbf{r}=(r_x,r_y)$ has images at positions $(r_x+mL+n\gamma L, r_y+nL)$, with $n$, $m$ integer and $\gamma=\dot\gamma t$ the total shear strain at time $t$.

\subsection{Pressure Tensor}

To determine the global rheology of the system we measure the pressure tensor of each configuration.  
We can break this pressure tensor into three pieces \cite{OHL,LeesEdwards}: the elastic part $\mathbf{p}^\mathrm{el}$, arising from the repulsive elastic forces of Eq.~(\ref{efel}),
\begin{equation}
\mathbf{p}^\mathrm{el}\equiv \frac{1}{L^2}\sum_{i<j}\mathbf{f}^\mathrm{el}_{ij}\otimes\mathbf{r}_{ij},
\label{epel}
\end{equation}
the dissipative part $\mathbf{p}^\mathrm{dis}$, arising from the dissipative forces of Eq.~(\ref{efdis}),
\begin{equation}
\mathbf{p}^\mathrm{dis}\equiv \frac{1}{L^2}\sum_{i<j}\mathbf{f}^\mathrm{dis}_{ij}\otimes\mathbf{r}_{ij},
\label{epdis}
\end{equation}
and the kinetic part $\mathbf{p}^\mathrm{kin}$ (sometimes called the streaming part),
\begin{equation}
\mathbf{p}^\mathrm{kin}\equiv\frac{1}{L^2}\sum_i m_i\delta\mathbf{v}_i\otimes\delta\mathbf{v}_i,
\label{epkin}
\end{equation}
where $\delta\mathbf{v}_i\equiv \mathbf{v}_i-\dot\gamma y_i\mathbf{\hat x}$ is the fluctuation away from the linear average velocity profile that characterizes the uniform shear strain flow.  The total pressure tensor is then,
\begin{equation}
\mathbf{p}=\mathbf{p}^\mathrm{el}+\mathbf{p}^\mathrm{dis}+\mathbf{p}^\mathrm{kin}.
\end{equation}

The average pressure $p$ and shear stress $\sigma$ in the system are then,

\begin{equation}
p=\frac{1}{2}\left[\langle p^\mathrm{}_{xx}\rangle +\langle p^\mathrm{}_{yy}\rangle\right],\quad
\sigma = -\langle p^\mathrm{}_{xy}\rangle,
\label{ep}
\end{equation}
where $\langle\dots\rangle$ represents an ensemble average over configurations in the sheared steady state.
Also of potential interest is the pressure anisotropy $\delta p$ and the deviatoric stress $\sigma_\mathrm{dev}$,
\begin{equation}
\delta p = \frac{1}{2}\left[\langle p_{xx}\rangle - \langle p_{yy}\rangle\right], \quad
\sigma_\mathrm{dev}=\sqrt{\delta p^2 + \sigma^2}.
\end{equation}
In the Appendix we present numerical results to show that while $\delta p$ can be non-negligible at low $\phi$ and low $Q$, the difference between $\sigma_\mathrm{dev}$ and $\sigma$ is always small for the range of parameters we consider.

Finally we can define the granular temperature $T_g$ in the usual way,
\begin{equation}
T_g \equiv \frac{1}{N}\sum_i m_i\langle |\delta\mathbf{v}_i|^2\rangle.
\end{equation}
We note that the kinetic part of the pressure $p^\mathrm{kin}$ is simply related to $T_g$ by $p^\mathrm{kin}=nT_g$ with $n=N/L^2$ the density of particles.

It is convenient to work in terms of dimensionless quantities.  We take the diameter of the small particles $d_s$, and the mass $m_0$, as our units of length and mass respectively.  We take $\tau_e$ as the unit of time.  With these choices, stress in two dimensions is measured in units of $m_0/\tau_e^2$, and so we can define a dimensionless pressure tensor $\mathbf{P}=(\tau_e^2/m_0)\mathbf{p}$.

Because we expect (and in the following section we confirm) that our system obeys Bagnoldian rheology, with $p,\sigma\sim\dot\gamma^2$ for sufficiently small $\dot\gamma$, we define the dimensionless Bagnold coefficients in terms of the components of $\mathbf{P}/(\dot\gamma\tau_e)^2$,
\begin{equation}
B_p\equiv \frac{p}{m_0\dot\gamma^2},\quad B_\sigma\equiv\frac{\sigma}{m_0\dot\gamma^2},
\end{equation}
and similarly for the separate pieces, $B_p^\mathrm{el}$, $B_p^\mathrm{dis}$, $B_p^\mathrm{kin}$, etc.
These dimensionless Bagnold coefficients are 
functions of only the dimensionless parameters $\phi$, $Q$, and $\dot\gamma\tau_e$.  As we will soon see, using $\tau_e$ as the unit of time will give Bagnold coefficients that become independent of $Q$ at small $Q$ for small $\dot\gamma\tau_e$ \cite{OlssonTeitelRheology}.

Note, the hard-core limit of infinitely stiff particles is usually considered as the limit $k_e\to\infty$, i.e. the interaction potential of Eq.~(\ref{eInteraction}) is so stiff that any particle overlaps are suppressed \cite{Campbell}.  By Eq.~(\ref{etaus}) this implies $\tau_e\to 0$ for particles with finite mass.
However, when expressed in the above dimensionless variables, we see that the hard-core limit is really the limit $\dot\gamma\tau_e\to 0$.  Thus, even for soft-core particles with finite $k_e$, and so finite $\tau_e$, we can reach the hard-core limit by taking a suitably small value of $\dot\gamma$ \cite{OHL}.  For sufficiently small $\dot\gamma\tau_e$ we expect the Bagnold coefficients $B_p$ and $B_\sigma$ to approach well defined values that depend on $\phi$ and $Q$, but are independent of $\dot\gamma\tau_e$.  These are the limiting hard-core values.  
How small $\dot\gamma\tau_e$ must be to reach this hard-core limit is not a priori known, it must be explicitly verified by simulations.
Note also that this hard-core limit places no constraint on the value of $Q$.  One should thus be careful to distinguish between the elasticity of particle interactions (i.e. stiffness of the particle core) governed by $k_e$ or equivalently $\tau_e$, and the elasticity of particle collisions (i.e. degree of energy conservation in a collision) governed by $Q$; the term {\em elasticity} has quite different meanings in these two different usages.  The behavior of the hard-core Bagnold coefficients, as a function of $\phi$ and $Q$, will be the main concern of this work.

\subsection{Simulation Method}

In  our numerical simulations,  we choose the diameter of the small particles to be $d_s=1$, and the mass $m_0=1$, and take the unit of time $\tau_e=1$ (which implies the elastic coupling $k_e=1$).
We integrate the equations of motion (\ref{eEqMotion}) using a modified velocity-Verlet algorithm with a Heun-like prestep to account for the velocity dependent acceleration.  We use an integration time step given by the following heuristic formula that varies according to the value of $Q$, $\Delta t/\tau_e = \min\{0.5/Q, 0.1, 0.2Q\}$.  The dependence of $\Delta t/\tau_e$ on $Q$ is motivated by the following physical picture: at large $Q$, particles move quickly so small time steps are needed to resolve all collisions; at very small $Q$ (large $k_d$), the dissipative force can become very large and too large a time step would cause particles to unphysically reverse direction rather than just slow down.  We have tested that our heuristic formula satisfactorily gives results independent of further decreasing the time step \cite{noteDt}.  

We simulate for a range of strain rates from $\dot\gamma\tau_e=10^{-3}$ down to $10^{-6}$.  
For $\dot\gamma\tau_e=10^{-5}$ (which corresponds to most of our presented results), we simulate out to a total strain $\gamma=\dot\gamma t$ of roughly $4<\gamma<100$, with the longest runs lying at intermediate values of $0.5\lesssim Q \lesssim 10$.  For $\dot\gamma\tau_e=10^{-6}$ we simulate to a total strain of roughly $0.12 <\gamma < 10$, again with the longest runs at intermediate values of $Q$.
In each case we exclude the initial 50\% of the run in order to reach steady state, and then collect data for our averages from the remainder of the run. For each parameter point $(\phi, Q, \dot\gamma\tau_e)$ we average over at least five independent runs.
Simulations at our largest $\dot\gamma$ are started from an initial random configuration at each $(\phi, Q)$; simulations at smaller $\dot\gamma$ start from a steady state configuration sampled from the simulation at the next larger $\dot\gamma$, at the same value of $(\phi,Q)$.

\section{Results}
\label{s3}

In this section we describe our numerical results.  We consider systems with a range of packing fractions from $\phi=0.60$ to $0.835$, and a range of $Q$ from $0.1$ to $500$.  Our range of $Q$ corresponds to a coefficient of restitution for two small particles ranging from $e=0$ to $0.9965$ (for $Q<0.564$, $e=0$; for $Q=2$, $e=0.3970$; for $Q=10$, $e=0.8373$).  In a previous work \cite{VOTBagnold1} we carried out a detailed critical scaling analysis of the jamming transition for the specific strongly inelastic case of $Q=1$, determining the value of the packing fraction at jamming to be $\phi_J=0.84335\pm 0.00005$.  Here we will present results to argue that the value of $\phi_J$, as well as all other critical parameters at jamming, are independent of the particular value of $Q$.  

Since our objective in the present work is to provide an understanding of the effect that varying $Q$ has on the rheology, rather than a quantitative analysis of critical behavior at jamming, our investigations will avoid getting too close to $\phi_J$; the closest we get to jamming will be $(\phi_J-0.835)/\phi_J=0.01$.  This allows us to work with the relatively small system size of $N=1024$ particles without incurring finite size effects, and relatively large strain rates $\dot\gamma\tau_e\ge 10^{-6}$ that still put us in the hard-core limit; this can be compared to the values $N=262144$ and $\dot\gamma\tau_e\ge 2\times 10^{-8}$ which we used in Ref.~[\onlinecite{VOTBagnold1}].

\subsection{Bagnold Coefficients}
\label{sBC}

\begin{figure}[h!]
\includegraphics[width=3.2in]{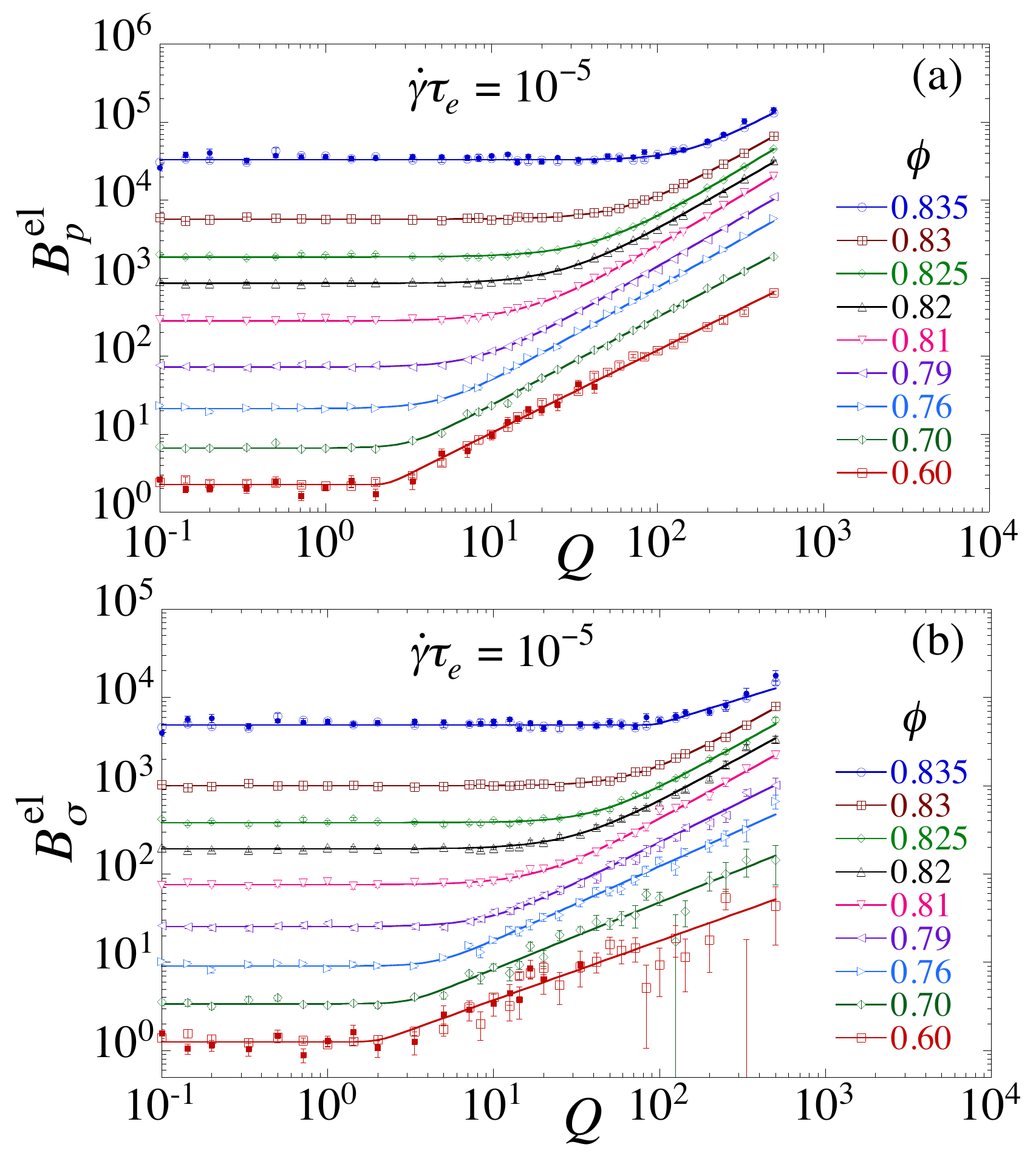}
\caption{(Color online) The elastic part of the Bagnold coefficients for (a) pressure $p$ and (b) shear stress $\sigma$ vs collision elasticity parameter $Q$ for different values of packing fraction $\phi$, which goes from 0.835 to 0.60 as the curves go from top to bottom.  Open symbols at all $\phi$ are for a shear strain rate $\dot\gamma\tau_e=10^{-5}$;  corresponding solid symbols at $\phi=0.60$ and $0.835$ are for $\dot\gamma\tau_e=10^{-6}$.  The absence of a dependence on $\dot\gamma\tau_e$ shows that results are in the hard-core limit.
}
\label{BpeBse-vs-Q}
\end{figure}

\begin{figure}[h!]
\includegraphics[width=3.2in]{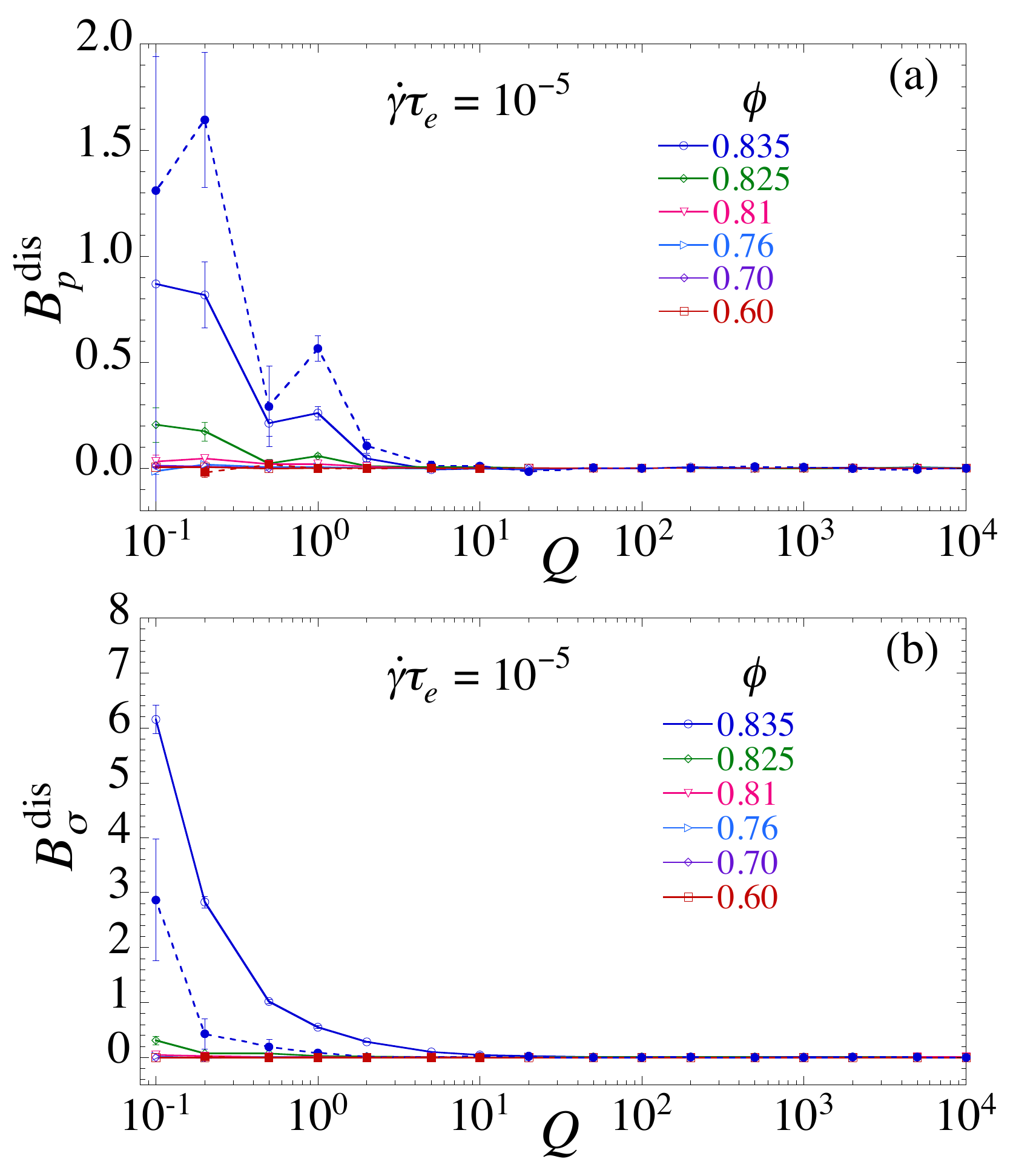}
\caption{(Color online) The dissipative part of the Bagnold coefficients for (a) pressure $p$ and (b) shear stress $\sigma$  vs collision elasticity parameter $Q$ for different values of packing fraction $\phi$.  Open symbols connected by solid lines at all $\phi$ are for a shear strain rate $\dot\gamma\tau_e=10^{-5}$;  corresponding solid symbols connected by dashed lines, at $\phi=0.60$ and $0.835$, are for $\dot\gamma\tau_e=10^{-6}$.  
}
\label{BpvBsv-vs-Q}
\end{figure}

\begin{figure}[h!]
\includegraphics[width=3.2in]{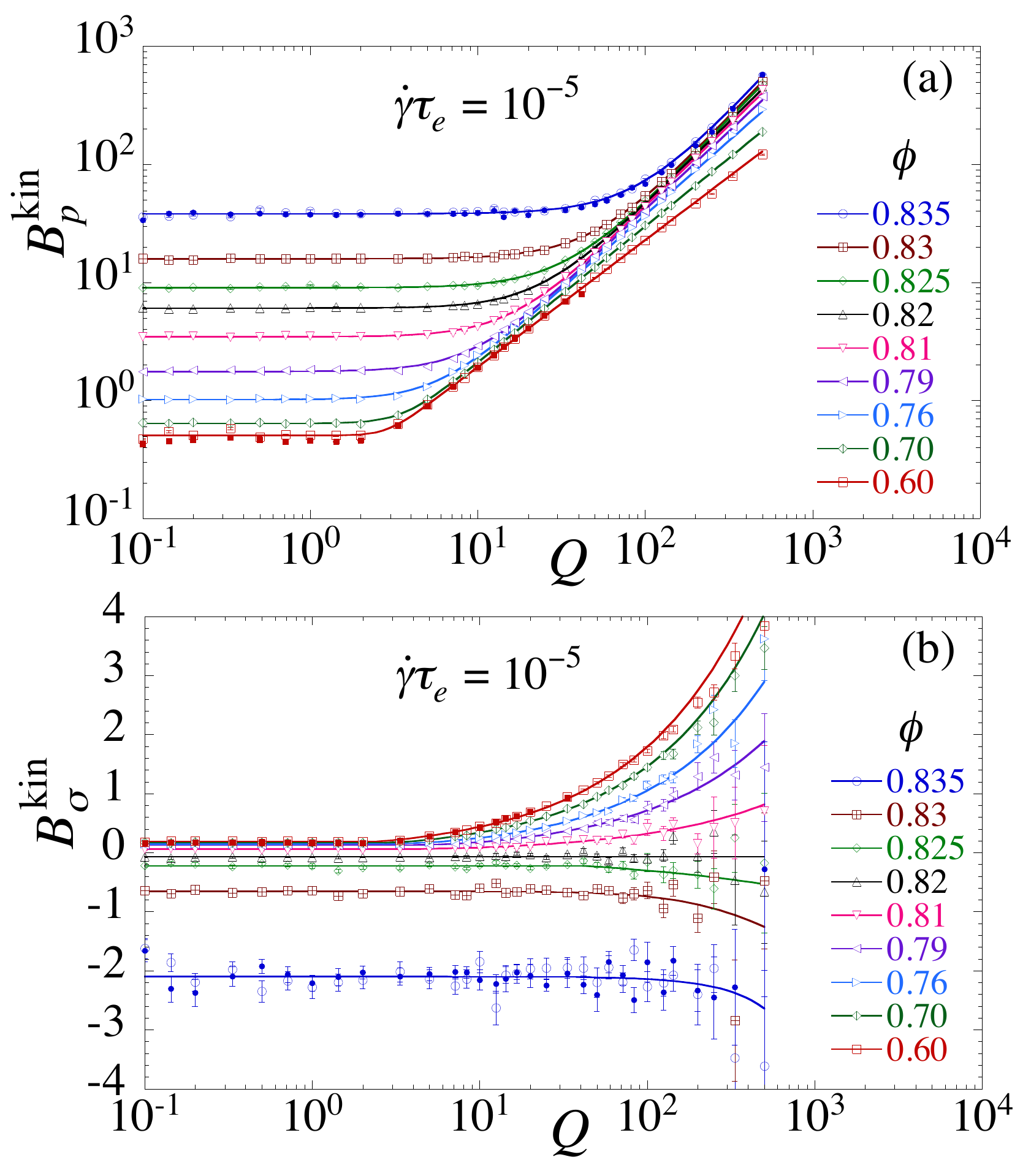}
\caption{(Color online) The kinetic part of the Bagnold coefficients for (a) pressure $p$ and (b) shear stress $\sigma$  vs collision elasticity parameter $Q$ for different values of packing fraction $\phi$.  In (a), $\phi$ decreases as the curves go from top to bottom; in (b) $\phi$ increases as the curves go from top to bottom.
Open symbols at all $\phi$ are for a shear strain rate $\dot\gamma\tau_e=10^{-5}$;  corresponding solid symbols at $\phi=0.60$ and $0.835$ are for $\dot\gamma\tau_e=10^{-6}$.  The absence of a dependence on $\dot\gamma\tau_e$ shows that results are in the hard-core limit.
Note the linear vertical scale in panel b, which is necessary since $B_\sigma^\mathrm{kin}$ changes sign.
}
\label{BpkBsk-vs-Q}
\end{figure}

In Figs.~\ref{BpeBse-vs-Q}, \ref{BpvBsv-vs-Q} and \ref{BpkBsk-vs-Q} we present our results for the elastic, dissipative, and kinetic parts of the Bagnold coefficients for pressure $p$ and shear stress $\sigma$, which we plot vs the elasticity parameter $Q$ for different fixed values of the packing fraction $\phi$.  We show results for a shear strain rate $\dot\gamma\tau_e=10^{-5}$, except for our smallest $\phi=0.60$ and largest $\phi=0.835$, where we show results for both $\dot\gamma\tau_e=10^{-5}$ (open symbols) and $\dot\gamma\tau_e=10^{-6}$ (solid symbols).  The observed absence of any dependence of the results on $\dot\gamma\tau_e$ (except for $B_p^\mathrm{dis}$ and $B_\sigma^\mathrm{dis}$ at the smallest $Q$ and largest $\phi$, see more below) indicates that our results are at sufficiently small $\dot\gamma\tau_e$ to represent the hard-core limit. If we wished to explore closer to the jamming point $\phi_J=0.84335$, it would be necessary to use smaller $\dot\gamma\tau_e$.

We consider first the dissipative parts $B_p^\mathrm{dis}$ and $B_\sigma^\mathrm{dis}$, shown in Fig.~\ref{BpvBsv-vs-Q}.  At small $Q$, we have found that the dissipative part fluctuates rapidly as a function of time, and so it was the most difficult of the three parts to compute accurately; our results here tend to be from longer runs than used elsewhere.  We see that both $B_p^\mathrm{dis}$ and $B_\sigma^\mathrm{dis}$ are essentially zero, except for the smallest $Q$ at the very largest $\phi$. For the largest $\phi=0.835$ we see that $B_\sigma^\mathrm{dis}$ decreases substantially as the strain rate decreases from $\dot\gamma\tau_e=10^{-5}$ (open circles) to $10^{-6}$ (solid circles).  Considering other values of $\dot\gamma\tau_e$ (not shown here) our results suggest that  $B_\sigma^\mathrm{dis}\sim\dot\gamma\tau_e$.  In contrast, $B_p^\mathrm{dis}$ at $\phi=0.835$ seems possibly to increase slightly as $\dot\gamma\tau_e$ decreases from $10^{-5}$ to $10^{-6}$; however, the estimated errors here are large and we cannot with confidence deduce a clear trend.  In any case, comparing Fig.~\ref{BpvBsv-vs-Q} with Fig.~\ref{BpeBse-vs-Q}, we see that, for all values of $\phi$ and $Q$ considered here, $B_p^\mathrm{dis}$ and $B_\sigma^\mathrm{dis}$ are completely negligible compared to $B_p^\mathrm{el}$ and $B_\sigma^\mathrm{el}$.  We therefore henceforth ignore these terms and take $B_{p,\sigma}=B_{p,\sigma}^\mathrm{el}+B_{p,\sigma}^\mathrm{kin}$. 

\begin{figure}[h!]
\includegraphics[width=3.2in]{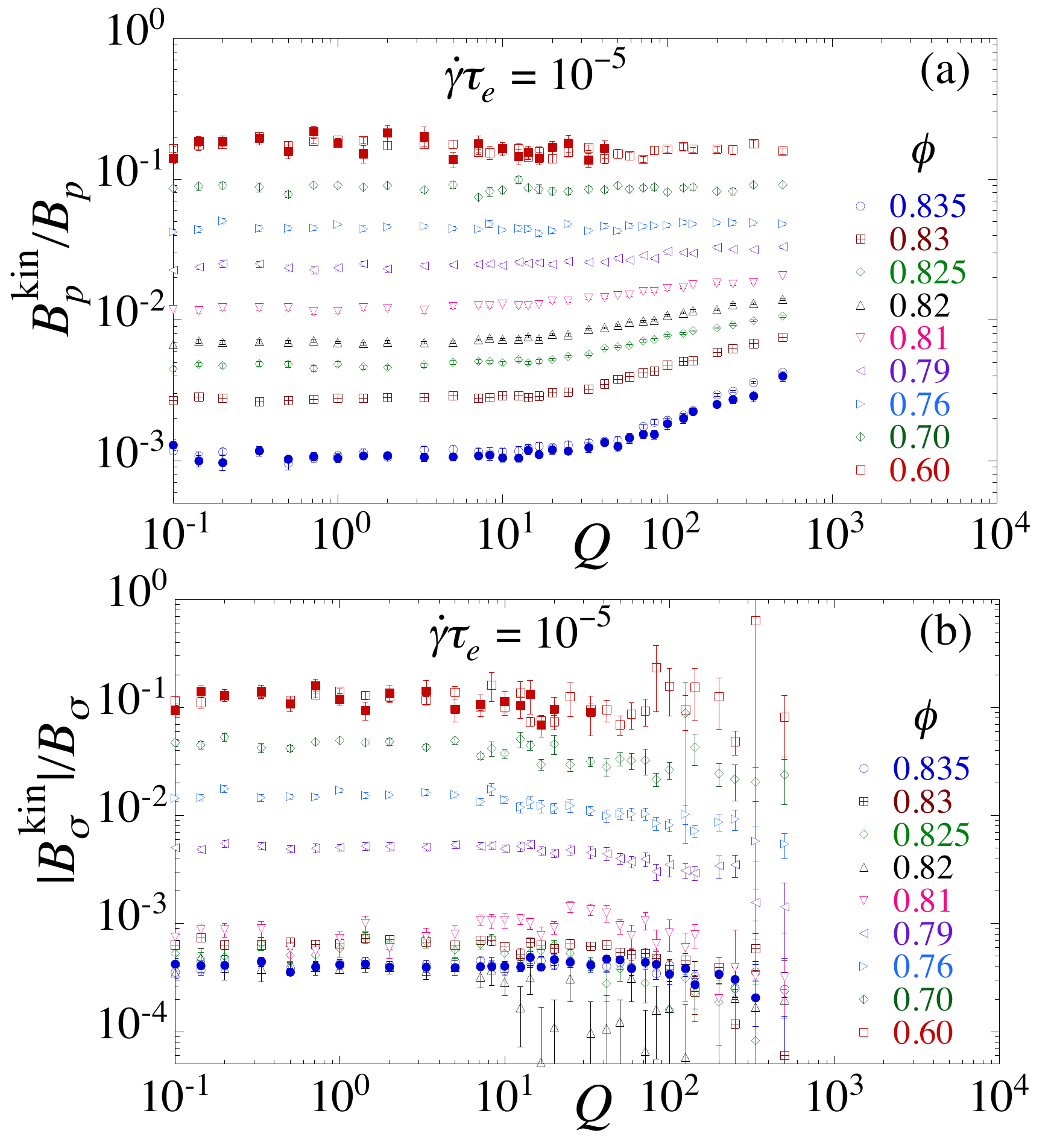}
\caption{(Color online) Relative contribution of the kinetic part to the total Bagnold coefficient:  (a) $B_p^\mathrm{kin}/B_p$ and (b) $|B_\sigma^\mathrm{kin}|/B_\sigma$ vs $Q$ for different fixed packing fraction $\phi$, which goes from 0.60 to 0.835 as the curves go from top to bottom. Open symbols at all $\phi$ are for a shear strain rate $\dot\gamma\tau_e=10^{-5}$;  solid symbols at $\phi=0.60$ and $0.835$ are for $\dot\gamma\tau_e=10^{-6}$.  Note, $B_p^\mathrm{kin}/B_p=nT_g/p$, with $T_g$ the granular temperature.
}
\label{RpkRsk-vs-Q}
\end{figure}

Considering next the kinetic parts $B_p^\mathrm{kin}$ and $B_\sigma^\mathrm{kin}$ in Fig.~\ref{BpkBsk-vs-Q} we see that as $\phi$ increases, $B_p^\mathrm{kin}$ steadily increases, while $B_\sigma^\mathrm{kin}$ decreases, becoming negative as $\phi$ gets close to the jamming $\phi_J= 0.84335$. 
In Fig.~\ref{RpkRsk-vs-Q} we plot the ratio $B_p^\mathrm{kin}/B_p$ and $|B_\sigma^\mathrm{kin}|/B_\sigma$ vs $Q$ for different fixed $\phi$.  We see that the relative contribution of the kinetic part to the total Bagnold coefficient is largest at our smallest $\phi$, where it is roughly 10\%.  But as $\phi$ increases, this relative contribution for $p$ drops rapidly to $0.1$--$0.5\%$ (depending on $Q$) at our largest $\phi=0.835$; for $\sigma$ it is in the range $0.05$--$0.1\%$.  Thus the contribution of the kinetic part becomes negligibly small as the jamming point is approached, justifying the neglect of this term in our earlier scaling analysis \cite{VOTBagnold1} of the divergence of $B_p$ and $B_\sigma$ at jamming for small $Q=1$.

We also note that, because of the relation between $p^\mathrm{kin}$ and the granular temperature $T_g$ ($p^\mathrm{kin}=nT_g$), we have $B_p^\mathrm{kin}/B_p=nT_g/p$.  
If our athermally sheared system was behaving the same as an equilibrium system at thermal temperature $T=T_g$, we would expect that, in the hard-core limit, $nT_g/p$ would be independent of the details of the dynamics and so a function solely of the packing fraction $\phi$, independent of the parameter $Q$.  The dependence of $B_p^\mathrm{kin}/B_p=nT_g/p$ on $Q$ observed in Fig.~\ref{RpkRsk-vs-Q}a, most notably at the larger values of $\phi$, thus indicates the difference between shear induced fluctuations and thermal fluctuations.


Finally we consider the elastic parts $B_p^\mathrm{el}$ and $B_\sigma^\mathrm{el}$ in Fig.~\ref{BpeBse-vs-Q}.  We see that at each $\phi$ there is a clear crossover value $Q^*(\phi)$, such that for $Q<Q^*$ the Bagnold coefficients are independent of $Q$, while for $Q>Q^*$ the Bagnold coefficients increase with $Q$ algebraically.  The value of $Q^*(\phi)$ increases as $\phi$ increases.  The same behavior is also observed in the kinetic parts $B_p^\mathrm{kin}$ and $B_\sigma^\mathrm{kin}$.  To determine the crossover values $Q^*$ we fit our data to the phenomenological form $B=C[1+(Q/Q^*)^{s}]^{q/s}$, which interpolates between the small and large $Q$ behaviors.  The exponent $q$ gives the large $Q$ algebraic behavior, while the parameter $s$ determines the sharpness of the crossover at $Q^*$.  The solid lines in Figs.~\ref{BpeBse-vs-Q} and \ref{BpkBsk-vs-Q} are the results of such fits.

\begin{figure}[h!]
\includegraphics[width=3.2in]{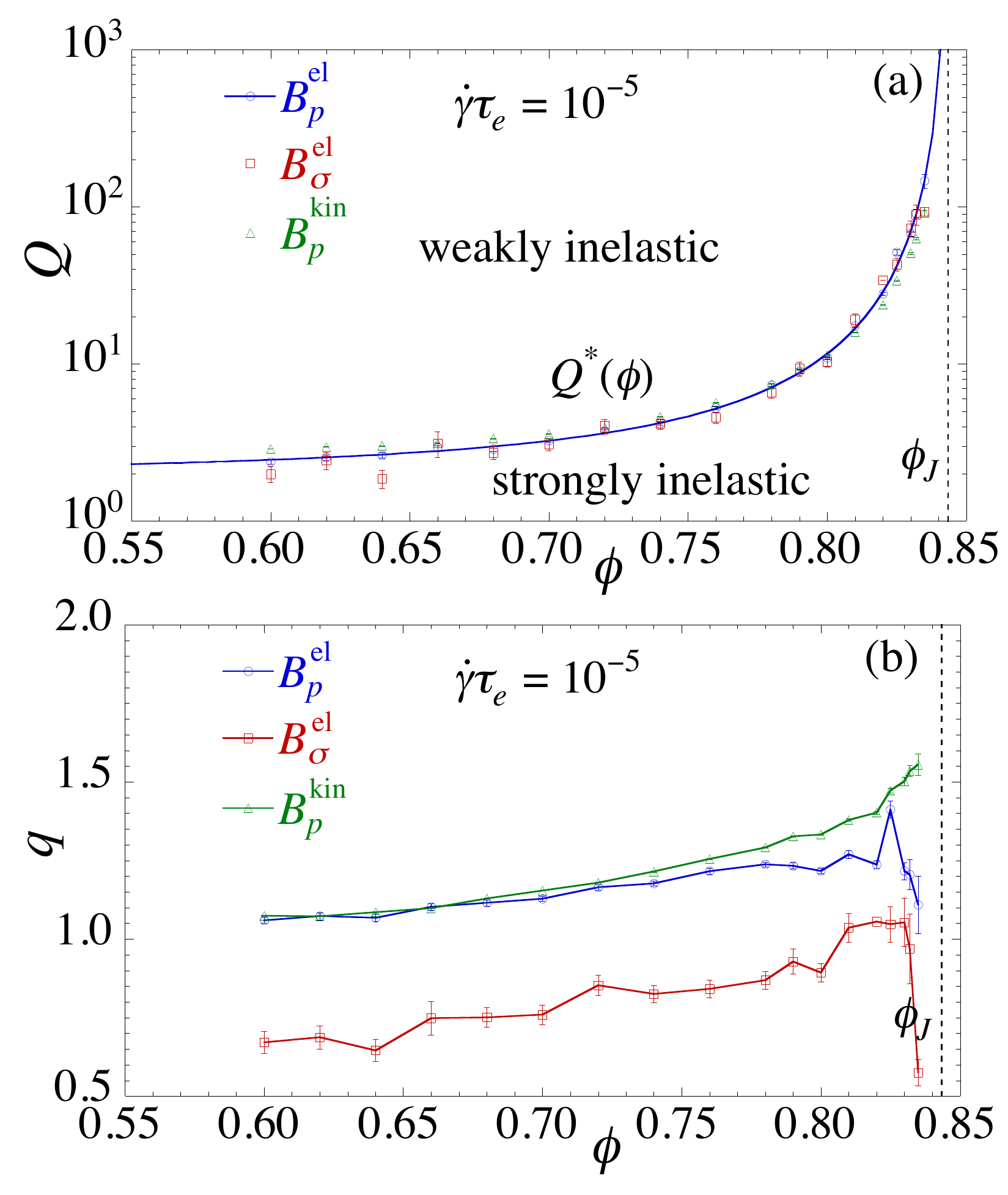}
\caption{(Color online) (a) Phase diagram in the $Q-\phi$ plane, showing the crossover $Q^*(\phi)$  that separates the region of strongly inelastic behavior from  weakly inelastic behavior.  We show values for $Q^*$ as determined independently from the Bagnold coefficients $B_p^\mathrm{el}$, $B_\sigma^\mathrm{el}$ and $B_p^\mathrm{kin}$ of Figs.~\ref{BpeBse-vs-Q} and \ref{BpkBsk-vs-Q}a; these are all found to agree. Solid line is a fit of $Q^*$, as obtained from $B_p^\mathrm{el}$, to the form $Q_0+c(\phi_J-\phi)^{-x}$ with fixed $\phi_J=0.84335$,  and yields the value $x\approx 1.65$. (b) Exponents $q$ that determine the large $Q$ algebraic increase of the Bagnold coefficients $B_p^\mathrm{el}$, $B_\sigma^\mathrm{el}$ and $B_p^\mathrm{kin}$.  In both panels the
vertical dashed line locates the jamming transition at $\phi_J=0.84335$.  Results are from simulations with shear strain rate $\dot\gamma\tau_e=10^{-5}$. 
}
\label{Qq-vs-phi}
\end{figure}

In Fig.~\ref{Qq-vs-phi}a we show the resulting phase diagram in the $Q-\phi$ plane, plotting the crossover $Q^*(\phi)$ that separates the region of strongly inelastic behavior ($Q<Q^*$) from weakly inelastic behavior ($Q>Q^*$).  We show $Q^*$ as determined from the above described fits, independently fitting to the data for $B_p^\mathrm{el}$, $B_\sigma^\mathrm{el}$ and $B_p^\mathrm{kin}$ shown previously in Figs.~\ref{BpeBse-vs-Q} and \ref{BpkBsk-vs-Q}a.  We see that the values of $Q^*$ obtained from these three quantities all agree nicely.  We do not show results for $B_\sigma^\mathrm{kin}$ since, as may be seen in Fig.~\ref{BpkBsk-vs-Q}b, the large scatter of the data at large $Q$, and the change in sign of $B_\sigma^\mathrm{kin}$ upon increasing $\phi$, gives a poor fit to our phenomenological form at the larger $\phi$.  In Fig.~\ref{Qq-vs-phi}b we show the fitted values of the exponent $q$ that give the large $Q$ algebraic growth in the Bagnold coefficients.  For the pressure parts,  $B_p^\mathrm{el}$ and $B_p^\mathrm{kin}$, we see that $q$ increases from roughly 1.1 to 1.5 as $\phi$ increases towards jamming; for the shear stress $B_\sigma^\mathrm{el}$, $q$ is noticeable smaller, increasing from roughly 0.6 to 1.0.  It is unclear if one should ascribe any fundamental significance to these particular values of $q$, of if they describe only empirical fits over the limited range of $Q$ we have investigated.

\subsection{The Shear Driven Jamming Transition}
\label{sJam}

We return to our results in Fig.~\ref{Qq-vs-phi}a.
We denote the region $Q<Q^*$, where the Bagnold coefficients become independent of $Q$, as the strongly inelastic region, while $Q>Q^*$ is the weakly inelastic region.  We discuss further some of the physical differences between these two regions in the next section.

An important feature of our result for $Q^*(\phi)$ is that $Q^*$ appears to be diverging as $\phi$ increases towards the jamming $\phi_J$.  This would imply that a system at any fixed value of $Q$ always crosses over from the weakly inelastic region into the strongly inelastic region, as $\phi$ increases above $\phi^*(Q)$, defined as the inverse of $Q^*(\phi)$.  Since jamming thus always takes place in the strongly inelastic region, and since in the strongly inelastic region the values of $B_p$ and $B_\sigma$ are independent of the particular value of $Q$, the asymptotic divergence of these quantities upon jamming is the same for all $Q$.  Hence the jamming packing fraction $\phi_J$, and all jamming critical exponents, are the same for all $Q$ and so equal to the values found in our earlier scaling analysis \cite{VOTBagnold1} carried out at the specific value of $Q=1$.  

Thus the only effect that increasing $Q$ has on the jamming transition is to decrease the region where strongly inelastic behavior (and its consequent critical scaling) holds.  As $Q$ diverges, and so collisions are perfectly elastic (energy conserving), this region shrinks to zero.   So it is only for this case of perfectly elastic collisions that the jamming critical behavior may become different.  The same conclusion was previously reached by OHL in Ref.~[\onlinecite{OHL}].

To support this conclusion, we fit our data for $Q^*$, as obtained from $B_p^\mathrm{el}$, to the form $Q^*(\phi)=Q_0+c(\phi_J-\phi)^{-x}$.  The solid line in Fig.~\ref{Qq-vs-phi}a is the result of such a fit keeping $\phi_J=0.84335$ fixed at the value determined by Ref.~[\onlinecite{VOTBagnold1}], and yields the exponent of divergence $x\approx 1.65\pm 0.02$ and $Q_0=1.89\pm0.06$.  If we instead let $\phi_J$ be a free parameter, then the fit gives $\phi_J=0.8425\pm  0.0010$, $x=1.59\pm0.07$ and $Q_0=1.8\pm0.1$, consistent with the previous result within the estimated errors.  The fitted values do not change significantly if we shrink the window of the fitted data closer to $\phi_J$.

\begin{figure}[h!]
\includegraphics[width=3.2in]{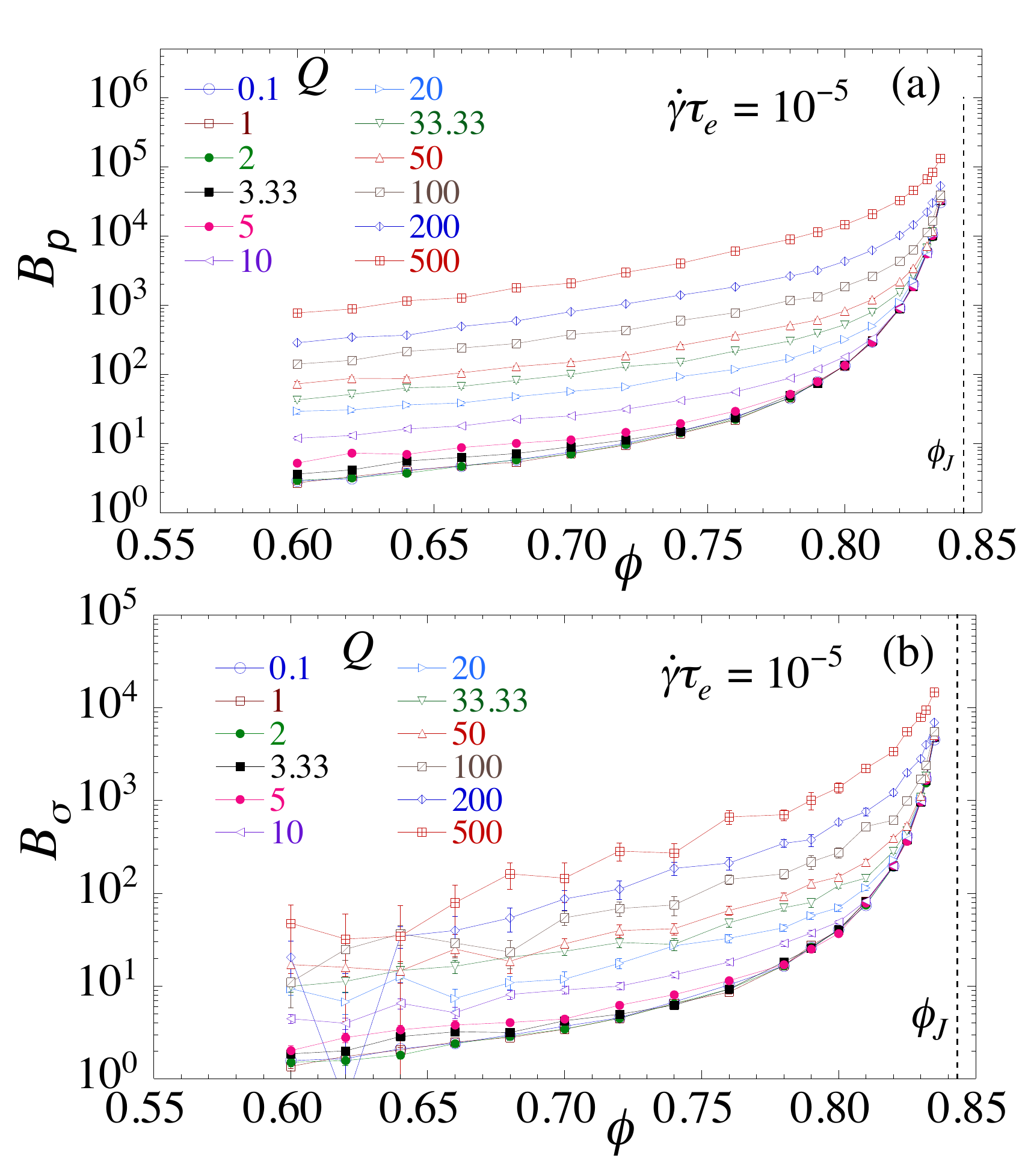}
\caption{(Color online) Total Bagnold coefficients (a) $B_p$ and (b) $B_\sigma$, for pressure $p$ and shear stress $\sigma$ respectively, vs packing fraction $\phi$, for different fixed values of the collision elasticity parameter $Q$; $Q$ increases from 0.1 to 500 as curves go from bottom to top.  The
vertical dashed line locates the jamming transition at $\phi_J=0.84335$.  Results are from simulations with shear strain rate $\dot\gamma\tau_e=10^{-5}$. 
}
\label{BpBs-vs-phi}
\end{figure}

To further illustrate the above point, in Fig.~\ref{BpBs-vs-phi} we plot the total Bagnold coefficients $B_p$ and $B_\sigma$ vs $\phi$, at different fixed values of $Q$.  We see that the curves for different $Q$ all are approaching a common curve, representing the strongly inelastic limit, as $\phi$ approaches $\phi_J$.  As $\phi$ decreases from $\phi_J$, the curves peel off from this common curve at a $\phi^*(Q)$ that decreases as $Q$ decreases.  For the several smallest values of $Q$, the curves overlap for the entire range of $\phi$ shown.

As $\phi$ approaches close to the jamming $\phi_J$, we expect to see a power law divergence of the Bagnold coefficients, $B_{p,\sigma}\sim (\phi_J-\phi)^{-\beta}$.  In our previous work of Ref.~[\onlinecite{VOTBagnold1}] at $Q=1$ we argued that to see the true asymptotic divergence of $B_p$ and $B_\sigma$ at jamming one needs to get extremely close to $\phi_J$ and use very small strain rates $\dot\gamma$.  Using a detailed critical scaling analysis, including leading corrections to scaling, we found $\beta= 5.0\pm0.4$ (see Ref.~\cite{VOTBagnold1} for a discussion of how this value of $\beta$ relates to those obtained in earlier numerical works).  We further showed that if one fits a simple power law to the Bagnold coefficients over a wider range of $\phi$ and $\dot\gamma$, one finds only an effective exponent $\beta_\mathrm{eff}<\beta$, whose value depends on the window of data used in the fit (see Fig.~7 of Ref.~[\onlinecite{VOTBagnold1}]).  
In the present work, we do not get anywhere close enough to the jamming critical point to see the true exponent $\beta$.  Nevertheless, we can still ask how the effective exponent $\beta_\mathrm{eff}$ will vary if one increases $Q$.

\begin{figure}[h!]
\includegraphics[width=3.2in]{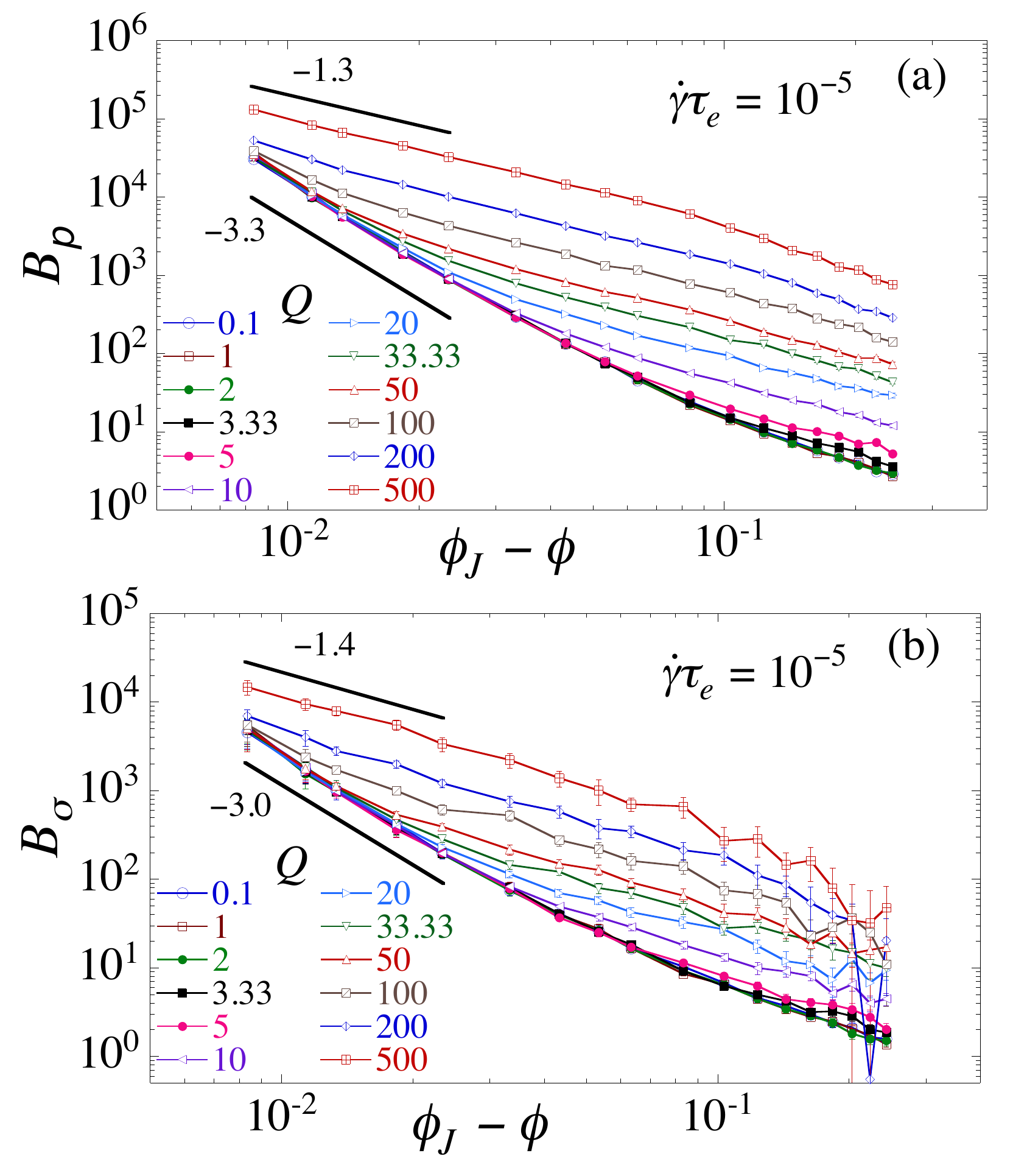}
\caption{(Color online) Total Bagnold coefficients (a) $B_p$ and (b) $B_\sigma$, for pressure $p$ and shear stress $\sigma$ respectively, vs $\phi_J-\phi$, for different fixed values of the collision elasticity parameter $Q$; $Q$ increases from 0.1 to 500 as curves go from bottom to top. We use $\phi_J=0.84335$ from Ref.~[\onlinecite{VOTBagnold1}].  The bold straight lines, with slopes as indicated in the figure, denote the approximate power law dependencies of our data closest to $\phi_J$, for the smallest and largest values of $Q$ (these are {\em not} the true power law divergences asymptotically close to $\phi_J$; see text).  Results are from simulations with shear strain rate $\dot\gamma\tau_e=10^{-5}$.  }
\label{BpBs-vs-dphi}
\end{figure}

In Fig.~\ref{BpBs-vs-dphi} we replot our data for $B_p$ and $B_\sigma$ vs $\phi_J-\phi$, using $\phi_J=0.84335$ from Ref.~[\onlinecite{VOTBagnold1}].  For small $Q$, where the data is in the strongly inelastic region for most of the values of $\phi$, we find for our range of data $\beta_\mathrm{eff}\approx 3.3$ for $B_p$ and $3.0$ for $B_\sigma$.  In contrast, for our largest $Q=500$, where most of the data remains in the weakly inelastic region, we find $\beta_\mathrm{eff}\approx 1.3$ for $B_p$ and $1.4$ for $B_\sigma$.  Thus $\beta_\mathrm{eff}$ can decrease substantially as $Q$ increases and collisions become increasingly elastic.   If we further allowed $\phi_J$ to be a free fitting parameter, rather than fixing it to its known value as we have done here, it is possible that yet other values of $\beta_\mathrm{eff}$ may be obtained.

We can, in principle, include the effects of a varying $Q$ within a critical scaling theory.  If we assume that for $Q<Q_0$ the Bagnold coefficients are independent of $Q$ for all $\phi$, then for $\delta Q\equiv Q-Q_0>0$ we can regard $\delta Q$ as a new scaling variable.  Since $\delta Q>0$ does not change the criticality of the jamming transition, it is an irrelevant variable, and thus has a negative scaling exponent.  We can then write the scaling equation \cite{VOTBagnold1} for $B_p$ as,
\begin{equation}
B_p(\phi, Q,\dot\gamma)=b^{\beta/\nu}f(\delta\phi b^{1/\nu}, \dot\gamma b^z, \delta Q b^{-x/\nu}, w b^{-\omega}),
\end{equation}
where $b$ is an arbitrary length rescaling factor, $\delta\phi=\phi_J-\phi$, $\nu$, and $z$ are the correlation length and dynamic critical exponents respectively, and $w$ is the leading irrelevant variable with exponent $\omega$.  If we then choose $b=\delta\phi^{-\nu}$, and consider the hard-core limit of $\dot\gamma\to 0$, the above becomes,
\begin{equation}
B_p(\phi,Q)=\delta\phi^{-\beta}f(1, 0, \delta Q\delta\phi^{x}, w \delta\phi^{\omega\nu}).
\label{Qscal}
\end{equation}
If we were close enough to the jamming point so that the leading irrelevant variable $w\delta\phi^{\omega\nu}$ could be ignored, then plotting $B_p \delta\phi^{\beta}$ vs $\delta Q\delta\phi^x$ would yield a collapse of the curves for different $\delta Q$.  However our prior work in Ref.~[\onlinecite{VOTBagnold1}] has shown that the leading irrelevant variable cannot be ignored for the range of parameters considered here, so such a collapse is not possible for our data.  
Nevertheless, Eq.~(\ref{Qscal}) still leads to the conclusion that the
crossover from the strongly inelastic limit  to the weakly inelastic limit is 
governed by the parameter $\delta Q\delta\phi^x$, and so takes place when $Q^*=Q_0+c\delta\phi^{-x}$, consistent with our numerical results in Fig.~\ref{Qq-vs-phi}a.

\begin{figure}[h!]
\includegraphics[width=3.2in]{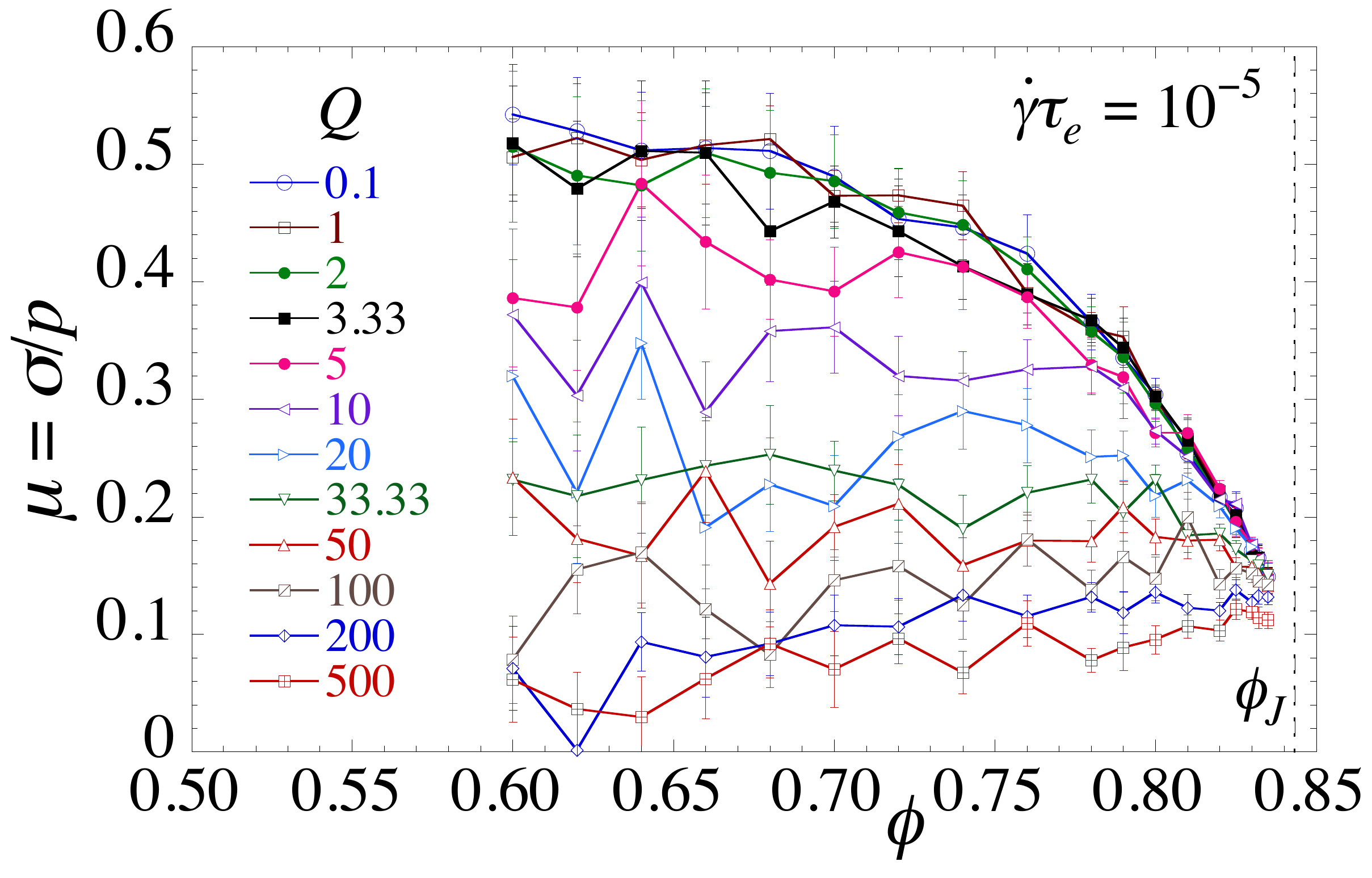}
\caption{(Color online) Macroscopic friction $\mu\equiv\sigma/p$ vs $\phi$ for different fixed values of $Q$; $Q$ increases from 0.1 to 500 as curves go from top to bottom.  The vertical dashed line locates the jamming transition at $\phi_J=0.84335$.  Results are from simulations with shear strain rate $\dot\gamma\tau_e=10^{-5}$.  }
\label{mu-vs-phi}
\end{figure}

Finally, we consider the macroscopic friction, $\mu\equiv\sigma/p$.  Although the individual particles have frictionless contacts, the macroscopic friction remains finite.  In Fig.~\ref{mu-vs-phi} we plot $\mu$ vs $\phi$ for different fixed values of $Q$.  We see that as $\phi$ approaches $\phi_J=0.84335$, $\mu$ approaches a common value $\mu_J$ for all $Q$.  In our prior work \cite{VOTBagnold1} we estimated $\mu_J\approx 0.093$.  Although our results for $\mu$ are rather noisy, the trend in behavior as $\phi$ and $Q$ are varied is clear.  At the smallest $Q$, the curves for $\mu$ overlap for all $\phi$, giving the limiting behavior of the strongly inelastic region, for which $\mu$ increases as $\phi$ decreases.  For larger $Q$, $\mu$ follows this common curve until $\phi$ decreases below $\phi^*(Q)$, at which point $\mu(\phi, Q)$ falls below the strongly inelastic limit.  For sufficiently large $Q$, $\mu$ even decreases as $\phi$ decreases, and can fall below the value of $\mu_J$.

\subsection{Strong vs Weak Inelastic Regions}

\begin{figure}[h!]
\includegraphics[width=3.2in]{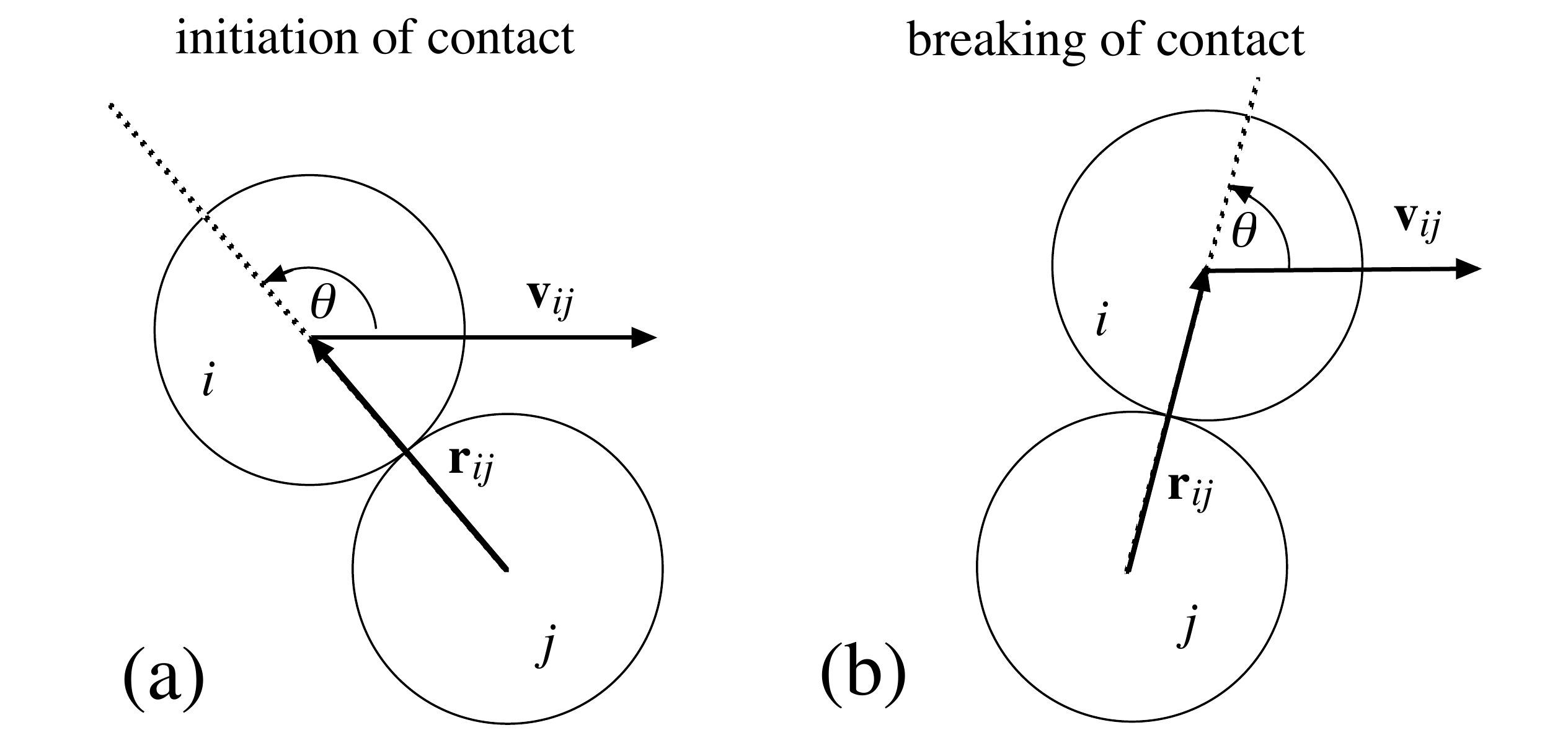}
\caption{Schematic of the collision of two particles $i$ and $j$. (a) Initiation of contact, where $90^\circ<\theta<270^\circ$, and (b) breaking of contact, where $-90^\circ<\theta<90^\circ$.  Here $\mathbf{r}_{ij}\equiv \mathbf{r}_i-\mathbf{r}_j$ and $\mathbf{v}_{ij}\equiv \mathbf{v}_i-\mathbf{v}_j$.
}
\label{collisions}
\end{figure}

Having found the crossover $Q^*(\phi)$ between the strongly and weakly inelastic regions, we can ask what different physical signatures characterize the behavior in the different regions.  One clear difference that we find concerns the angle of collision impact.  To measure this, let us define,
\begin{equation}
\mathbf{r}_{ij}\equiv \mathbf{r}_i-\mathbf{r}_j,\quad \mathbf{v}_{ij}\equiv\mathbf{v}_i-\mathbf{v}_j,
\end{equation}
as the position and velocity of particle $i$ with respect to particle $j$.  We then define the angle $\theta$ as the angle by which one must rotate $\mathbf{v}_{ij}$ to align it parallel with $\mathbf{r}_{ij}$ \cite{thetaNote}.  For two particles just initiating a contact, we must have $\mathbf{\hat v}_{ij}\cdot\mathbf{\hat r}_{ij} =\cos\theta< 0$, so that the particles are driven into each other, as illustrated in Fig.~\ref{collisions}a.  In this case we must have $90^\circ<\theta<270^\circ$.  For two particles just breaking a contact, we must have $\mathbf{\hat v}_{ij}\cdot\mathbf{\hat r}_{ij} =\cos\theta> 0$, so that the particles are driven away from each other, as illustrated in Fig.~\ref{collisions}b.  In this case we must have $-90^\circ <\theta<90^\circ$.  

\begin{figure}[h!]
\includegraphics[width=3.2in]{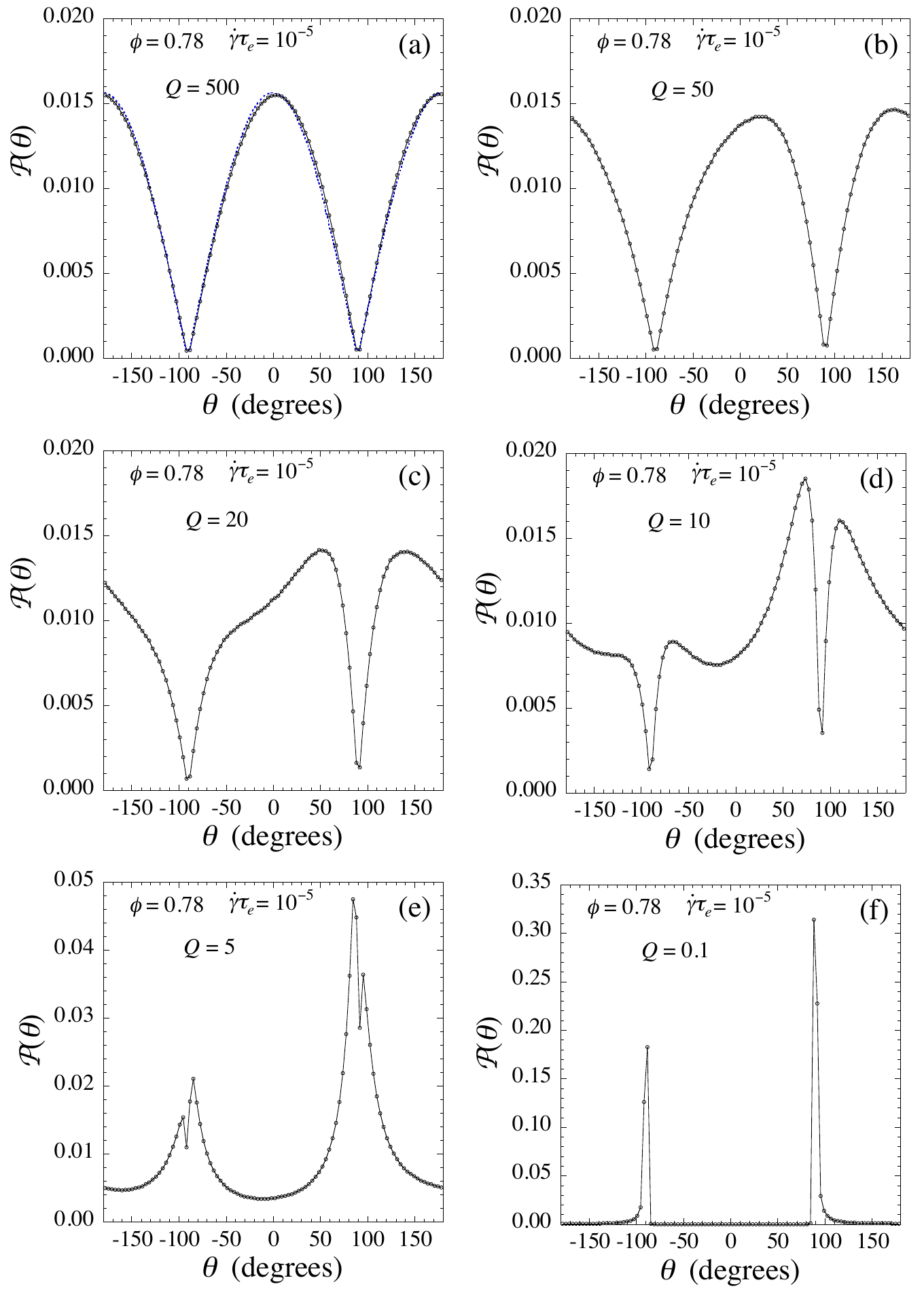}
\caption{(Color online) Histograms ${\cal P}(\theta)$ vs collision angle $\theta$, at initiation and breaking of particle contacts (see Fig.~\ref{collisions} for definition of $\theta$), at $\phi=0.78$.  Results are shown for different values of the collision elasticity parameter $Q$. (a) $Q=500$, (b) $Q=50$, (c) $Q=20$, (d) $Q=10$, (e) $Q=5$, and (f) $Q=0.1$.  The crossover $Q^*\approx 7.35$ at this value of $\phi$.  Dotted blue line in (a) is a fit to ${\cal P}(\theta)=C|\cos(\theta)|$.
Results are for the shear strain rate $\dot\gamma\tau_e=10^{-5}$.}
\label{Ptheta}
\end{figure}

Measuring the value of $\theta$ each time a contact is initiated and each time a contact is broken, we construct a histogram ${\cal P}(\theta)$ which combines both contact initiation and contact breaking events.  In Fig.~\ref{Ptheta} we plot ${\cal P}(\theta)$ vs $\theta$ at several different values of $Q$, for the particular case of $\phi=0.78$ for which $Q^*\approx 7.35$.  For the weakly inelastic case of $Q=500\gg Q^*$ in Fig.~\ref{Ptheta}a, we see that ${\cal P}(\theta)\sim|\cos\theta|$, as would be expected if the collision impact parameter, $b=|\mathbf{r}_{ij}\times\mathbf{\hat v}_{ij}|$,  is distributed uniformly on the interval $-d_{ij} < b < d_{ij}$.  Thus, deep in the weakly inelastic region collisions occur at all angles, with a normal head-on collision at $\theta=0$ being the most likely.  In contrast, for the strongly inelastic case of $Q=0.1\ll Q^*$ in Fig.~\ref{Ptheta}f, we see that ${\cal P}(\theta)$ has sharp peaks at $\theta=\pm 90^\circ$, and ${\cal P}(\theta)$ is a minimum at $\theta=0$.  Thus, in the strongly inelastic region collisions involve mostly tangential relative motion between particles.  Figures~\ref{Ptheta}b--e  show ${\cal P}(\theta)$ at intermediate value of $Q$ to illustrate how the distribution transforms between these two limits.  We observe similar behavior at other values of $\phi$.  
The reason for this behavior is simple.  As $Q$ gets small, the dissipative force of Eq.~(\ref{efdis}) damps out the relative normal motion of particles in contact, but does not effect the relative tangential motion.

\begin{figure}[h!]
\includegraphics[width=3.2in]{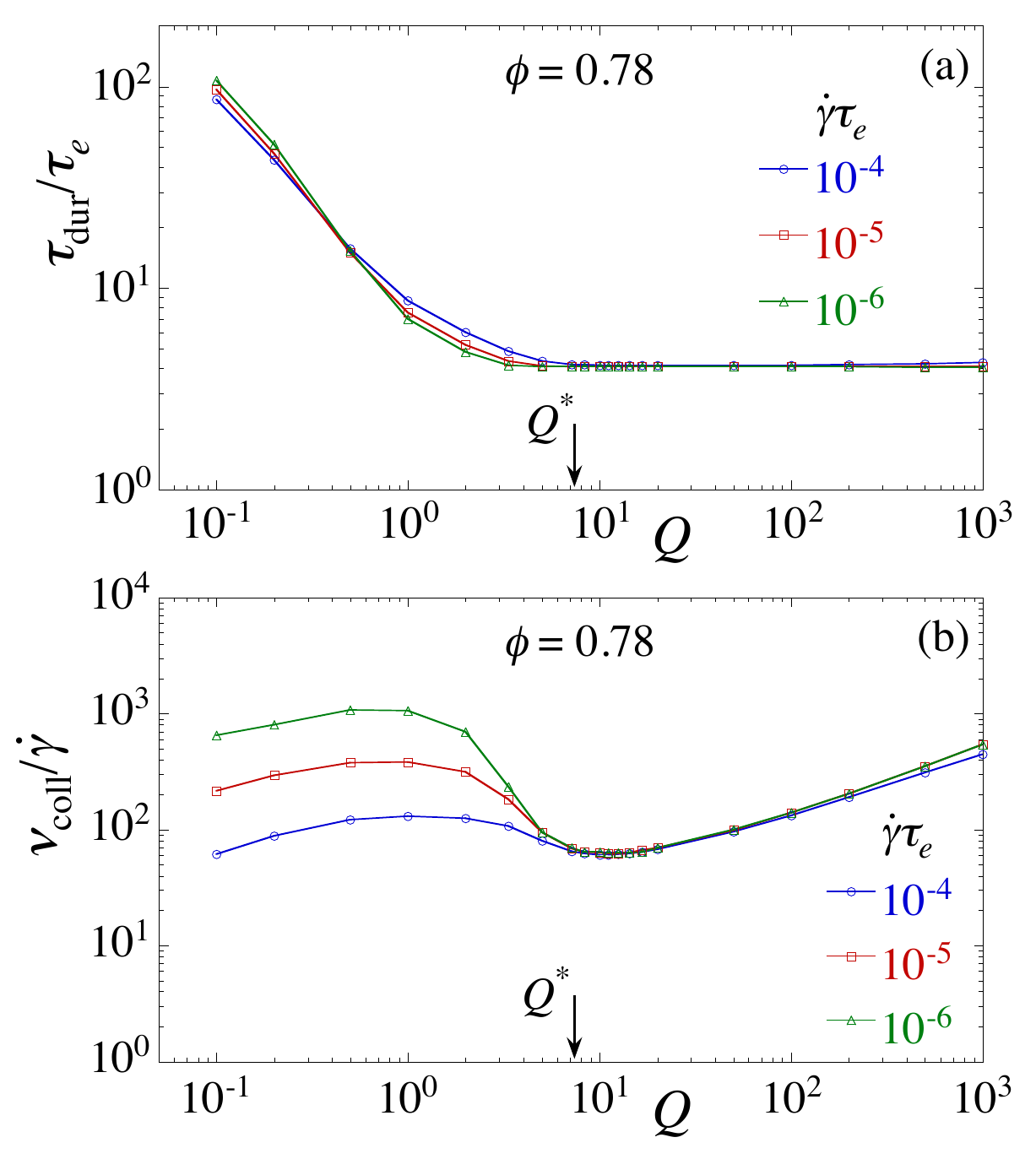}
\caption{(Color online) (a) Average duration of collision $\tau_\mathrm{dur}/\tau_e$, from time of initiation of contact to time of breaking of contact, vs collision elasticity parameter $Q$.  (b) Ratio of average collision rate to strain rate, $\nu_\mathrm{coll}/\dot\gamma$, vs $Q$.  In both panels results are shown for packing fraction $\phi=0.78$ and shear strain rates $\dot\gamma\tau_e=10^{-4}$, $10^{-5}$, and $10^{-6}$. The crossover $Q^*\approx 7.35$, separating strongly from weakly inelastic regions, is denoted by the vertical arrow.
}
\label{tdur}
\end{figure}

We can get further insight into the different nature of collisions in the strong vs weak inelastic regions by considering the average time duration of a collision, $\tau_\mathrm{dur}$, and the average collision rate, $\nu_\mathrm{coll}$;  $\tau_\mathrm{dur}$ is defined as the time from the initiation of a particular particle contact to the breaking of that contact, $\nu_\mathrm{coll}$ is defined as the average number of collisions per unit time divided by the number of particles.
In  Fig.~\ref{tdur}a we plot $\tau_\mathrm{dur}/\tau_e$ vs $Q$, for the particular case of $\phi=0.78$ and several different strain rates $\dot\gamma\tau_e$.  We see that $\tau_\mathrm{dur}/\tau_e$ is essentially constant in the weakly inelastic region $Q>Q^*$; this constant value $\tau_\mathrm{dur}/\tau_e\approx 4$  is just slightly bigger than the large $Q$ value for an isolated head-on collision between a small and big particle, which is 3.84 \cite{Schafer}.  But  as $Q$ decreases into the strongly inelastic region, we see that $\tau_\mathrm{dur}/\tau_e$ rises over two orders of magnitude. For the strain rates considered here, we see that $\tau_\mathrm{dur}/\tau_e$ varies little with $\dot\gamma\tau_e$.

In Fig.~\ref{tdur}b we plot the dimensionless $\nu_\mathrm{coll}/\dot\gamma$ vs $Q$ for the same parameters as in Fig.~\ref{tdur}a, i.e. $\phi=0.78$ and $\dot\gamma\tau_e=10^{-4}$, $10^{-5}$ and $10^{-6}$.  We see that in the weakly inlastic region, $Q>Q^*$, the curves for different $\dot\gamma\tau_e$ coincide, showing that  $\nu_\mathrm{coll}\propto \dot\gamma$.  In the strongly inelastic region, $Q<Q^*$, however, the curves separate, with the smaller strain rate curve lying above the higher strain rate curve; this shows that in the strongly inelastic region the collision rate $\nu_\mathrm{coll}$ grows more slowly than linearly with increasing $\dot\gamma$.

\begin{figure}[h!]
\includegraphics[width=3.2in]{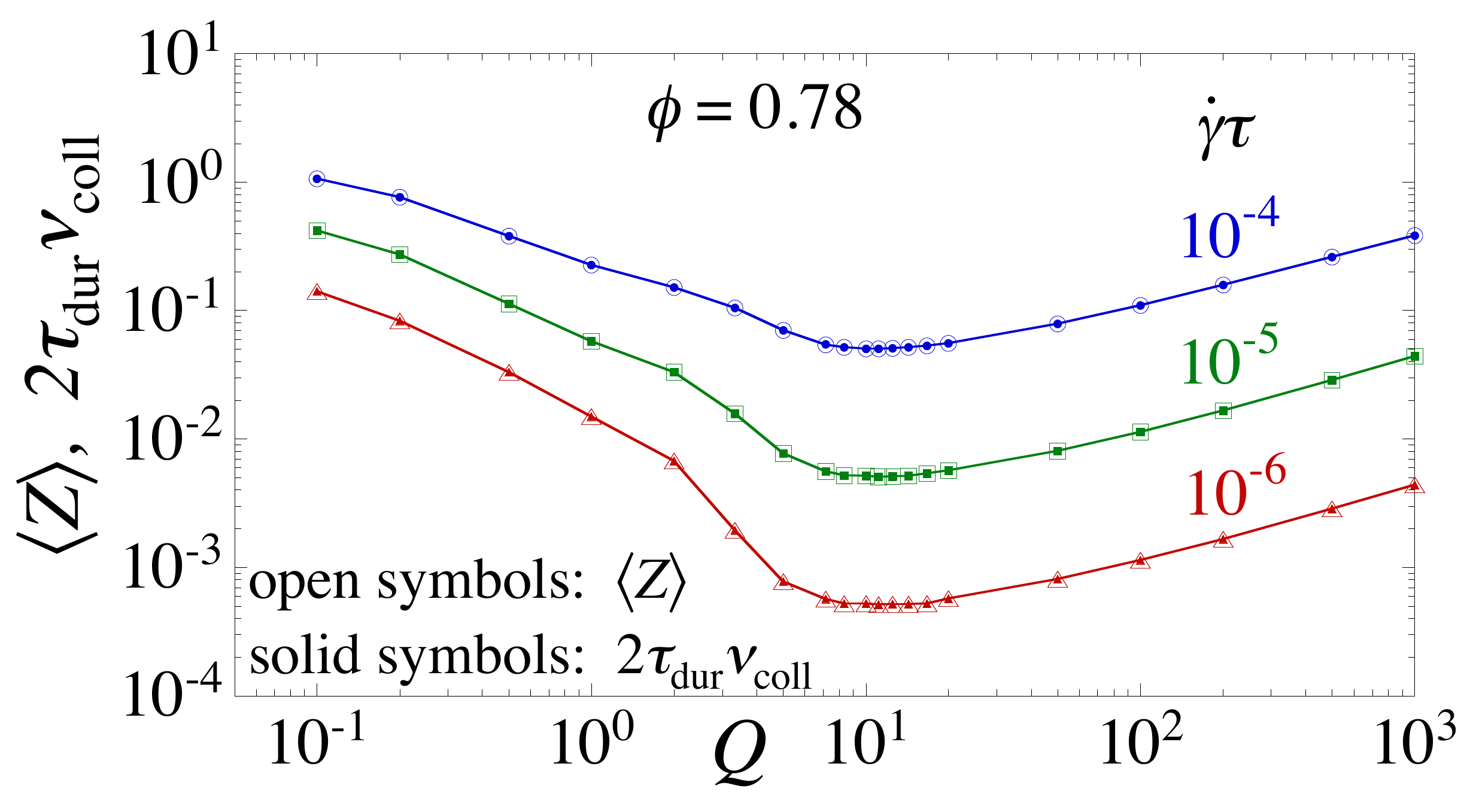}
\caption{(Color online) Comparison of average particle contact number $\langle Z\rangle$ (open symbols) with $2\tau_\mathrm{dur}\nu_\mathrm{coll}$ (solid symbols) vs $Q$ at packing fraction $\phi=0.78$ for strain rates $\dot\gamma\tau_e=10^{-4}$, $10^{-5}$ and $10^{-6}$.  
}
\label{z-2tdurnucoll}
\end{figure}

In Ref.~\cite{OHL} OHL give a relation between 
the average instantaneous particle contact number $\langle Z\rangle$ and the collision duration $\tau_\mathrm{dur}$ and rate $\nu_\mathrm{coll}$.  $Z$ is the number of contacts a given particle has with the other particles at any particular instant in time.   They argue that $\langle Z\rangle = 2 \tau_\mathrm{dur}\nu_\mathrm{coll}$.  Using our data in Fig.~\ref{tdur} we find excellent agreement with this prediction \cite{OHLnote2}, as we show in Fig.~\ref{z-2tdurnucoll}.  From this relation we can infer the behavior of $\langle Z\rangle$ as a function of the strain rate $\dot\gamma$.
As argued by OHL \cite{OHL}, and reported by us recently \cite{OlssonTeitelRheology}, we find that in all regions below $\phi_J$,  $\langle Z\rangle\to 0$ as $\dot\gamma\to 0$.  However, as we show now,  the manner in which $\langle Z\rangle$ vanishes with decreasing $\dot\gamma$ differs in the two regions.  In the weakly inelastic region, $Q>Q^*$, since from Fig.~\ref{tdur} we see that both $\tau_\mathrm{dur}/\tau_e$ and $\nu_\mathrm{coll}/\dot\gamma$ are independent of the strain rate $\dot\gamma\tau_e$, we conclude that $\langle Z\rangle \propto \dot\gamma\tau_e$ as $\dot\gamma\to 0$.  But in the strongly inelastic region, $Q<Q^*$, we see that $\tau_\mathrm{dur}/\tau_e$ is roughly independent of $\dot\gamma\tau_e$ but $\nu_\mathrm{coll}/\dot\gamma$ is decreasing more slowly than linearly in the strain rate; hence we conclude that in the strongly inelastic region $\langle Z\rangle$ decreases more slowly than linearly with $\dot\gamma\tau_e$ as $\dot\gamma\to 0$.

\begin{figure}[h!]
\includegraphics[width=3.2in]{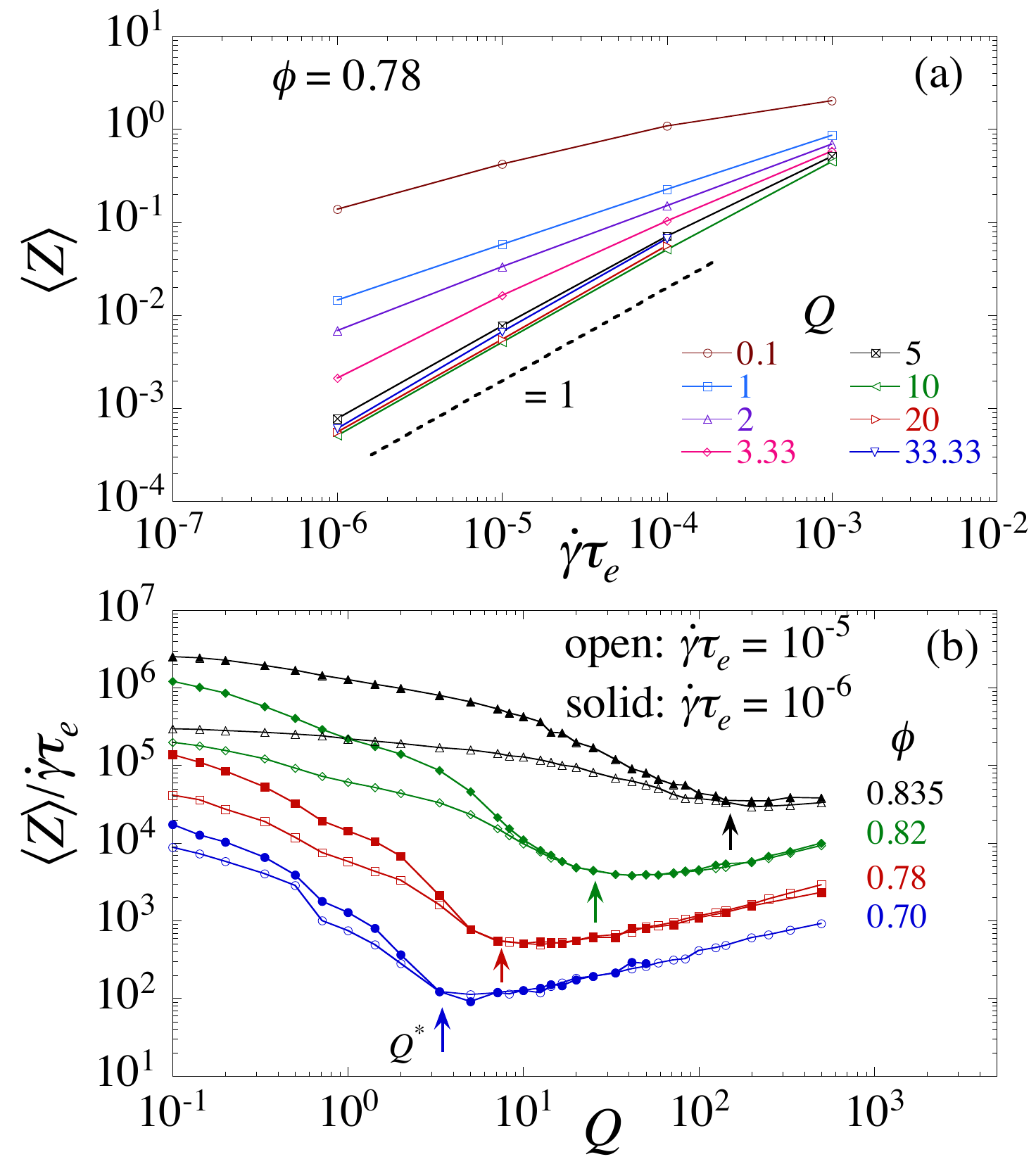}
\caption{(Color online) (a) Average contact number $\langle Z\rangle$ vs strain rate $\dot\gamma\tau_e$ at $\phi=0.78$, for different values of $Q$.  The dashed line has slope of unity, indicating a linear relation at large $Q$. (b) $\langle Z\rangle/\dot\gamma\tau_e$ vs $Q$ for $\dot\gamma\tau_e=10^{-5}$ (open symbols) and $10^{-6}$ (solid symbols) at several different values of $\phi$.  Vertical arrows indicate the location of the crossover $Q^*$ at each different $\phi$.
}
\label{z}
\end{figure}

We show this explicitly in Fig.~\ref{z}.  In Fig.~\ref{z}a we show $\langle Z\rangle$ vs $\dot\gamma\tau_e$, for different values of $Q$, at the packing fraction $\phi=0.78$ where $Q^*\approx 7.35$.  We see that for large $Q\gtrsim Q^*$, $\langle Z\rangle$ decreases linearly with $\dot\gamma\tau_e$ as $\dot\gamma\tau_e\to 0$.  However for $Q<Q^*$, $\langle Z\rangle$ decreases more slowly as $\dot\gamma\tau_e\to 0$.   In Fig.~\ref{z}b we plot $\langle Z\rangle/\dot\gamma\tau_e$ vs $Q$ for several different values of $\phi$, at the two strain rates $\dot\gamma\tau_e=10^{-5}$ and $10^{-6}$.  For $Q>Q^*$ we see that $\langle Z\rangle/\dot\gamma\tau_e$ is independent of $\dot\gamma\tau_e$, thus confirming that $\langle Z\rangle\propto \dot\gamma\tau_e$. For $Q<Q^*$, however, the curves separate, with the smaller $\dot\gamma\tau_e=10^{-6}$ curve lying above the $\dot\gamma\tau_e=10^{-5}$ curve; this indicates that $\langle Z\rangle$ is decreasing less rapidly than $\dot\gamma\tau_e$, as implied by the behavior of $\nu_\mathrm{coll}$ in Fig.~\ref{tdur}b.  We also see that $\langle Z\rangle$ is non-monotonic in $Q$. This is a reflection of the increase in $\nu_\mathrm{coll}$ with increasing $Q$ at large $Q$, and the increase in $\tau_\mathrm{dur}$ with decreasing $Q$ at small $Q$.

One can ask what is the mechanism by which increasing the packing fraction $\phi$ results in an increase in the threshold $Q^*$ below which normal relative motion is damped out, and the contact number $\langle Z\rangle$ decreases more slowly with $\dot\gamma\tau_e$.  Our preliminary investigation into this question suggests the following picture:  for $Q>Q^*$ essentially all collisions are isolated binary collisions, where only two particles are in contact during any given collision; for $Q<Q^*$ however, we find that collisions become correlated, so that many collisions involve multiple particles in mutual contact.  The number of such mutually contacted particles is found to grow as the density $\phi$ increases.  Such an effect is presumably related to the decreasing free volume available to the particles as $\phi$ increases, and serves to renormalize the dissipative mechanism for damping relative normal motion, which leads to the increasing $Q^*$ as $\phi$ increases.  We leave further detailed exploration of this effect to future work.

As a final note, 
we have denoted the $Q<Q^*$ region, where the dissipative coupling $k_d$ is large, as ``strongly inelastic" (and $Q>Q^*$ as ``weakly inelastic") in analogy to the behavior of an isolated colliding pair.  This analogy is supported by our results in Fig.~\ref{tdur}a, where we see that the collision duration time $\tau_\mathrm{dur}$ is small and constant for $Q>Q^*$, but grows rapidly once $Q$ decreases below $Q^*$.  But this nomenclature is perhaps misleading in one respect.  The rate of energy dissipation per volume in the system is $\Gamma=\sigma\dot\gamma = m_0 B_\sigma\dot\gamma^3$.  From Fig.~\ref{BpeBse-vs-Q}b or Fig.~\ref{BpBs-vs-phi}b we see that $B_\sigma$, and hence $\Gamma$, increases as $Q$ increases into the weakly inelastic region.  Thus dissipation is {\em smaller} in the strongly inelastic region than it is in the weakly inelastic region, contrary to what one might naively expect.  The reason for this behavior is given by Fig.~\ref{Ptheta}.  In the region $Q<Q^*$ the many particle steady state arranges itself so that collisions tend to involve only tangential relative motion. Since the dissipative force of Eq.~(\ref{efdis}) depends only on the difference of the normal components of the particles' velocities, little energy is dissipated in such collisions.  Our terminology ``strongly inelastic" for $Q<Q^*$ thus refers specifically to the effect of a collision on the {\em normal component} of the relative motion of the colliding particles; tangential relative motion remains undamped at any $Q$.

\subsection{Granular Constitutive Equations}

In the previous sections we have discussed the dependence of quantities on the packing fraction $\phi$, as appropriate for systems at constant volume.  In the literature on hard granular materials, where pressure is often the regarded as the control parameter rather than volume, it is common to express quantities as a function of the {\it inertial number} $I$ \cite{daCruz, Pouliquen,Forterre,Lemaitre,Roux},
\begin{equation}
I\equiv \dfrac{\dot\gamma}{\sqrt{p/m_0}}=\dfrac{1}{\sqrt{B_p}},
\label{eIB}
\end{equation}
rather than the packing fraction $\phi$.
Since in the hard-core limit $B_p$ is independent of $\dot\gamma$ and depends only on $\phi$ and $Q$,
we have $I(\phi, Q)$, which can be inverted to write as $\phi(I,Q)$. We thus can regard $I$ rather than $\phi$ as the control parameter; thus in the hard-core limit, $I$ is independent of the separate values of $\dot\gamma$ and $p$, and depends only on the combination as above in Eq.~(\ref{eIB}). Moreover, since in the hard-core limit $B_\sigma$ is also a function of only $\phi$ and $Q$, we can substitute for $\phi$ in terms of $I$ and write $B_\sigma(I,Q)$.  We thus get the macroscopic friction $\mu=\sigma/p=B_\sigma/B_p$ as a function of $I$ and $Q$.  The two functions $\phi(I,Q)$ and $\mu(I,Q)$ are known as the {\it constitutive equations}.
The jamming point corresponds to $I\to 0$ (i.e. $B_p\to\infty$).  

For sufficiently small $I$ close to jamming, it is observed empirically that the functions $\phi(I,Q)$ and $\mu(I,Q)$ can be written in the following form, 
\begin{equation}
\phi(I)=\phi_J-c_\phi I^a,\qquad \mu(I)=\mu_J +c_\mu I^b.
\label{phimu}
\end{equation}
At the level of an empirical result, the coefficients $c_\phi$ and $c_\mu$ and exponents $a$ and $b$ might depend on $Q$; however we will argue below that as $I\to 0$, these parameters are in fact independent of $Q$.

It is often argued \cite{Pouliquen, Forterre, Lemaitre} that $\phi$ and $\mu$ are linear in the inertial number $I$, i.e. $a=b=1$, for small $I$.  However the evidence for such linear behavior seems to be best found in systems in which there is a microscopic inter-particle friction \cite{daCruz,Bouzid}.  For frictionless particles, such as we consider here, Peyneau and Roux \cite{Roux} considered a strongly inelastic system and found, from fits to a range $10^{-5}\le I\le 10^{-2}$, the exponents $a\approx b\approx 0.4$. Earlier work by da~Cruz et al. \cite{daCruz} similarly found $\mu$ to be sublinear  in $I$ at small $I$ for frictionless particles.  Later work by Bouzid et al. \cite {Bouzid} claimed $b=1/2$, based on fits to a range $4\times 10^{-4}\le I\le 10^{-1}$, for strongly inelastic frictionless particles.

In Ref.~\cite{VOTBagnold1} we have shown that, in the asymptotic limit $I\to 0$, the form of the constitutive equations of Eq.~(\ref{phimu}) follows directly from the algebraic divergence of $B_p$ and $B_\sigma$ as $\phi\to\phi_J$, and the exponents $a$ and $b$ of the constitutive equations are related to the exponents $\beta$ and $\omega\nu$ of Eq.~(\ref{Qscal}) by,
\begin{equation}
a=2/\beta,\qquad b=2\omega\nu/\beta=\omega\nu a.
\label{ab}
\end{equation}
In Ref.~\cite{VOTBagnold1} we found $\omega\nu\approx 1$ and $\beta\approx 5$, thus suggesting $a=b=2/\beta\approx 0.4$, in agreement with Peyneau and Roux \cite{Roux}.  Since we have argued in Sec.~\ref{sJam} that the jamming transition always takes place within the strongly inelastic region $\phi>\phi^*(Q)$, where behavior is independent of the parameter $Q$, this then implies that the constitutive equations (\ref{phimu}) likewise must be independent of $Q$, for sufficiently small $I$; hence we conclude that all the parameters that appear in Eq.~(20) are independent of $Q$ as $I\to 0$.

\begin{figure}[h!]
\includegraphics[width=3.2in]{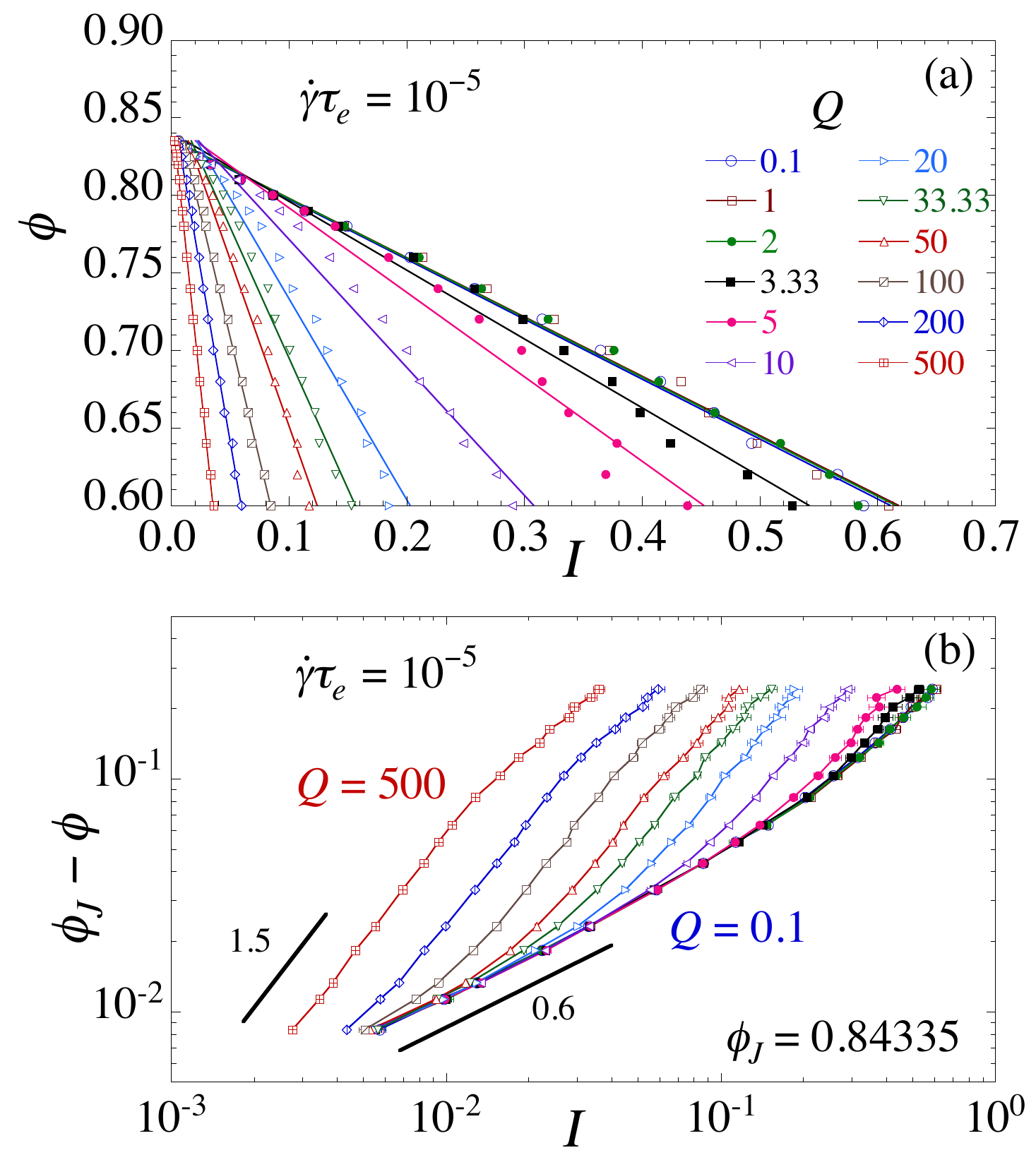}
\caption{(Color online) (a) Packing fraction $\phi$ vs inertial number $I=\dot\gamma/\sqrt{p/m_0}$ for various values of $Q$.  Straight lines are linear fits to the data.  (b) Same as panel (a) but plotted as $\phi-\phi_J$ vs $I$ on a log-log scale, where $\phi_J=0.84335$ is taken from Ref.~\cite{VOTBagnold1}.  Solid lines correspond to power law relations $\phi_J-\phi\propto I^{a_\mathrm{eff}}$ with $a_\mathrm{eff}=1.5$ and $0.6$ as shown.  Symbols in panel (b) correspond to the legend in panel (a).
Results are for a strain rate of $\dot\gamma\tau_e=10^{-5}$.
}
\label{phi-vs-I}
\end{figure}

The above discussion was concerned with behavior asymptotically close to the jamming point $I\to 0$.  It is interesting to now consider how the functions $\phi(I,Q)$ and $\mu(I,Q)$ behave as $I$ increases out of the asymptotic small $I$ region where Eq.~(\ref{phimu}) holds, and in particular when the system crosses into the weakly inelastic region $\phi<\phi^*(Q)$.
In Fig.~\ref{phi-vs-I}a we plot packing fraction $\phi$ vs inertial number $I$ for various values of $Q$ at a strain rate $\dot\gamma\tau_e=10^{-5}$.  From our results in Sec.~\ref{sBC} we know this $\dot\gamma\tau_e$ is small enough to put one in the hard-core limit for the range of parameters considered here.  On the linear-linear scale of \ref{phi-vs-I}a, the data look qualitatively like the results of da Cruz et al. \cite{daCruz}, and at moderate to high values of $Q$ the data appear well approximated by a linear fit (the solid lines in the figure) over the wide range of $I$ shown.  But if one looks closely at the data at the smallest $I$, approaching $\phi_J$, one finds that these linear fits are really not doing very well.  We see this explicitly in Fig.~\ref{phi-vs-I}b, where we plot $\phi_J-\phi$ vs $I$ on a log-log scale; we use $\phi_J=0.84335$ from our earlier work in Ref.~\cite{VOTBagnold1}.  We see that the slopes of the data at small $I$ are not in general equal to unity, the value expected if we had the exponent $a=1$.  Fig.~\ref{phi-vs-I}b is just the analog of Fig.~\ref{BpBs-vs-dphi}a, and as found there, the curves at different $Q$ all approach a common curve, characteristic of the strongly inelastic region $\phi>\phi^*(Q)$, as one gets sufficiently close to the jamming point $I\to 0$.

We may try to empirically fit our small $I$ data in Fig.~\ref{phi-vs-I}b to the form of Eq.~(\ref{phimu}). But since our data is not sufficiently close to the asymptotic $I\to 0$ limit, rather than finding the true asymptotic critical exponent $a$
we will find for each $Q$ only an {\em effective} power law exponent $a_\mathrm{eff}$, that depends both on the value of $Q$ and the range of $I$ used in the fit.  We see from Fig.~\ref{phi-vs-I}b that, for our range of data, this $a_\mathrm{eff}$ ranges from about 0.6 at our smallest $Q$ to 1.5 at our largest $Q$.
That we find  $a_\mathrm{eff}\approx 0.6$ at the smallest $Q$, rather than the value 0.4 expected by our work in Ref.~\cite{VOTBagnold1} and as found by Peyneau and Roux \cite{Roux}, is simply because our small $Q$ data, though already in the strongly inelastic region, is not at sufficiently small $I$ to be in the true asymptotic jamming critical region.
We thus see that, as with $\beta_\mathrm{eff}$ of Fig.~\ref{BpBs-vs-dphi}a, the value of $a_\mathrm{eff}$ for a finite range of $I$ can be strongly affected by the value of $Q$.

\begin{figure}[h!]
\includegraphics[width=3.2in]{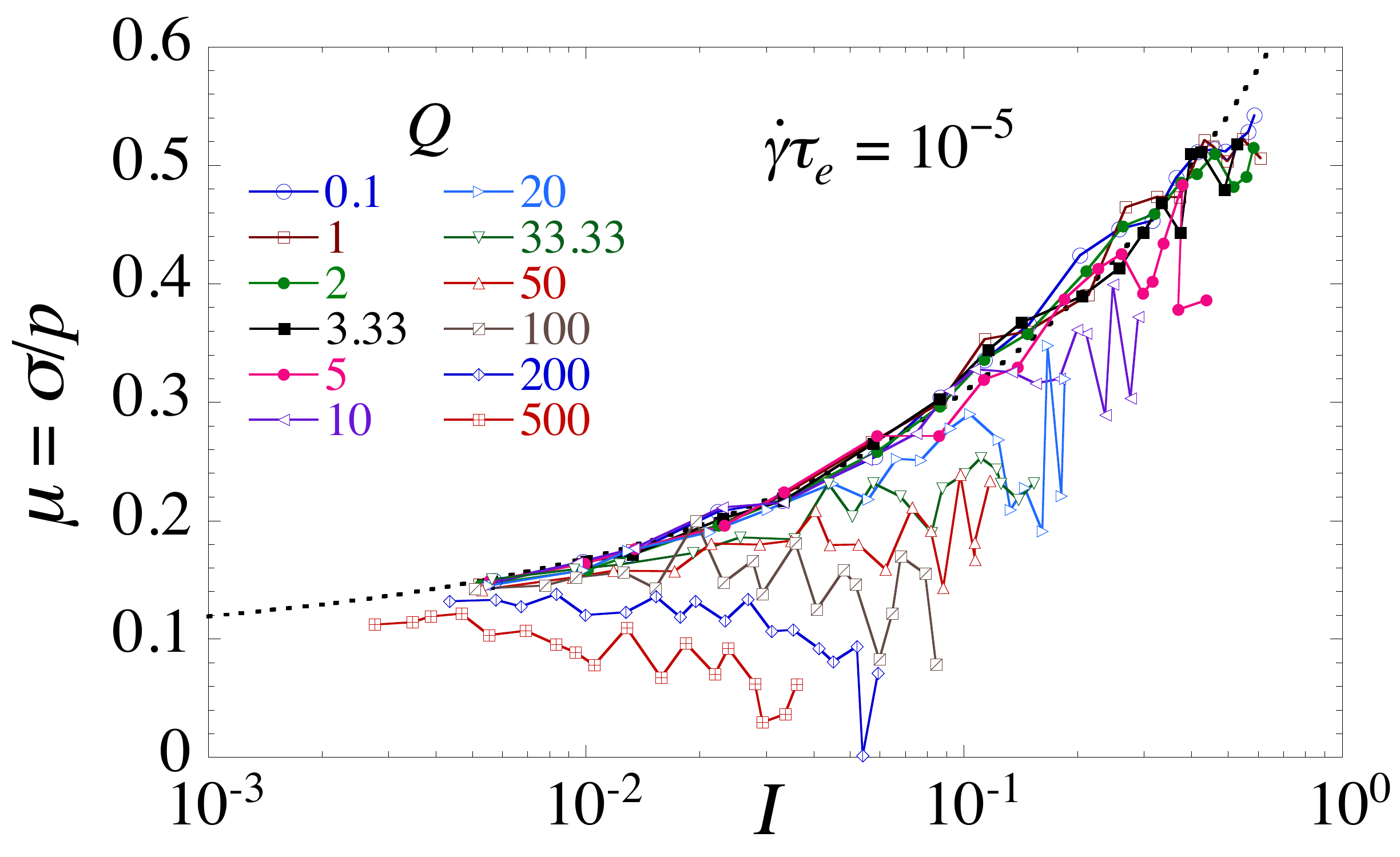}
\caption{(Color online) Macroscopic friction $\mu=\sigma/p$ vs the inertial number $I=\dot\gamma/\sqrt{p/m_0}$ for various $Q$.
Results are for a strain rate of $\dot\gamma\tau_e=10^{-5}$. 
The dashed line is a fit to the form $\mu=\mu_J+c_\mu I^{b_\mathrm{eff}}$, with fixed $\mu_J=0.093$ from Ref.~\cite{VOTBagnold1}, and gives the value $b_\mathrm{eff}\approx 0.46$.
}
\label{mu-vs-I}
\end{figure}

Finally we consider the macroscopic friction $\mu=\sigma/p$.  In Fig.~\ref{mu-vs-I} we plot $\mu$ vs $I$ for different $Q$ at the strain rate $\dot\gamma\tau_e=10^{-5}$.   Again we see that curves for different $Q$ approach a common curve characteristic of the strongly inelastic region $\phi>\phi^*(Q)$, as one gets close to jamming, $I\to 0$.  But as $I$ increases, the curves peel away from this common curve at an $I^*(Q)$ that decreases as $Q$ increases.  Similar results were found by Lois et al. \cite{Lois2}.  Fig.~\ref{mu-vs-I} is just the analog of Fig.~\ref{mu-vs-phi}, and again we see that for large $Q$, $\mu$ can decrease below the value $\mu_J$ at jamming  as $I$ increases.  Fitting our data for the smallest $Q=0.1$ in Fig.~\ref{mu-vs-I} to the form of Eq.~\ref{phimu}, and taking $\mu_J=0.093$ from Ref.~\cite{VOTBagnold1},  we find the exponent $b_\mathrm{eff}= 0.46\pm 0.02$.  This is larger than the expected $b\approx 0.4$ in the asymptotic  limit $I\to 0$ \cite{VOTBagnold1,Roux}, but close to the value $1/2$ found by Bouzid et al. \cite{Bouzid}.  As with $\beta_\mathrm{eff}$ and $a_\mathrm{eff}$, the value of $b_\mathrm{eff}$ depends on the range of $I$ over which one fits, and may be influenced by the value of $Q$ if part of the fitted data lies outside the strongly inelastic region.

\section{Conclusions}
\label{s4}
We have investigated the shear driven Bagnold rheology of a simple model of athermal, soft-core, bidisperse, frictionless disks in two dimensions, as a function of the packing fraction $\phi$ and a parameter $Q$ that controls the elasticity of collisions.
We have shown that there is a $Q^*(\phi)$ that marks a  sharp, but non-singular, crossover from a region 
characteristic of strongly inelastic collisions ($Q<Q^*$), where normal relative motion of particles is strongly damped and the resulting relative motion is mostly tangential, to a region characteristic of weakly inelastic collisions ($Q>Q^*$).  
In the strongly inelastic region, transport coefficients are independent of the value of $Q$, while in the weakly inelastic region, transport coefficients grow algebraically with increasing $Q$.
We have presented evidence that $Q^*(\phi)$ diverges as $\phi\to\phi_J$, the jamming transition, thus arguing that sufficiently close to $\phi_J$ one is always in the strongly inelastic region.  As a consequence, the value of $\phi_J$, and the critical exponents that characterize the divergence of the Bagnold transport coefficients, do not depend on the value of $Q$.  However, we have also shown that {\em effective} exponents, obtained from fitting over windows of data wider than the true asymptotic region close to $\phi_J$, can vary depending on the width of the data window and the value of $Q$.

We  have shown that the weakly inelastic region is characterized by a collision rate $\nu_\mathrm{coll}$ and an average particle contact number $\langle Z\rangle$ that scale linearly with the strain rate $\dot\gamma$, while the duration time of collisions $\tau_\mathrm{dur}$ is largely independent of $Q$ and $\dot\gamma$.  Deep in the weakly inelastic region (i.e. nearly elastic), collisions are uniformly distributed over all impact parameters, and particles tend to bounce off each other after they collide.

In the strongly inelastic region, the collision rate $\nu_\mathrm{coll}$ and contact number $\langle Z\rangle$ still vanish as $\dot\gamma\to 0$, but they decrease more slowly than linearly in $\dot\gamma$.  As $Q$ decreases into the strongly inelastic region, the collision duration time $\tau_\mathrm{dur}$ grows rapidly, and collisions increasingly involve tangential relative motion between particles.

We believe that this crossover to tangential relative motion as $Q$ decreases is a result of two different effects: (i) the damping out of particles' relative motion in the normal direction due to the dissipative force of Eq.~(\ref{efdis}), and (ii) the decreasing free volume available for particle motion as the packing fraction $\phi$ increases; this also greatly restricts relative motion in the normal direction, but less so for tangential relative motion.  We believe it is this second effect that is responsible for the divergence of $Q^*$ as $\phi\to\phi_J$.

We have also examined the macroscopic friction $\mu$ in our model and find that, while $\mu$ in the strongly inelastic region increases as $\phi$ decreases (or as inertial number $I$ increases), once one enters the weakly inelastic region $\mu$ can decrease as $\phi$ further decreases (or as $I$ further increase) and even fall below the value $\mu_J$ at jamming.

To summarize, we have shown that while the critical behavior asymptotically close to jamming is always characteristic of the strongly inelastic region, and so independent of the elasticity of collisions $Q$, the effect of collision elasticity can be clearly seen  as one moves away from jamming.

\section*{Acknowledgements}

This work was supported by National Science Foundation Grant No. DMR-1205800, the Swedish Research Council Grant No. 2010-3725, and the European Research Council under the European UnionÕs Seventh Framework Programme (FP7/2007-2013), ERC Grant Agreement No. 306845. Simulations were performed on resources provided by the Swedish National Infrastructure for Computing (SNIC) at PDC and HPC2N.  We wish to thank H. Hayakawa and M. Otsuki for helpful discussions. 

\section*{Appendix}

In this Appendix we provide numerical results for the anisotropy in pressure,
\begin{equation}
\delta p \equiv \frac{1}{2}\left[ \langle p_{xx}\rangle - \langle p_{yy}\rangle \right],
\end{equation}
and the deviatoric stress,
\begin{equation}
\sigma_\mathrm{dev}\equiv\sqrt{\delta p^2 +\sigma^2},
\end{equation}
where $\sigma=-\langle p_{xy}\rangle$.  The eigenvalues of the stress tensor are just $p\pm\sigma_\mathrm{dev}$, so a finite $\delta p$ results in a slight shift in the orientation of the principle axes of the stress tensor from those of the strain tensor.

\begin{figure}[h!]
\includegraphics[width=3.2in]{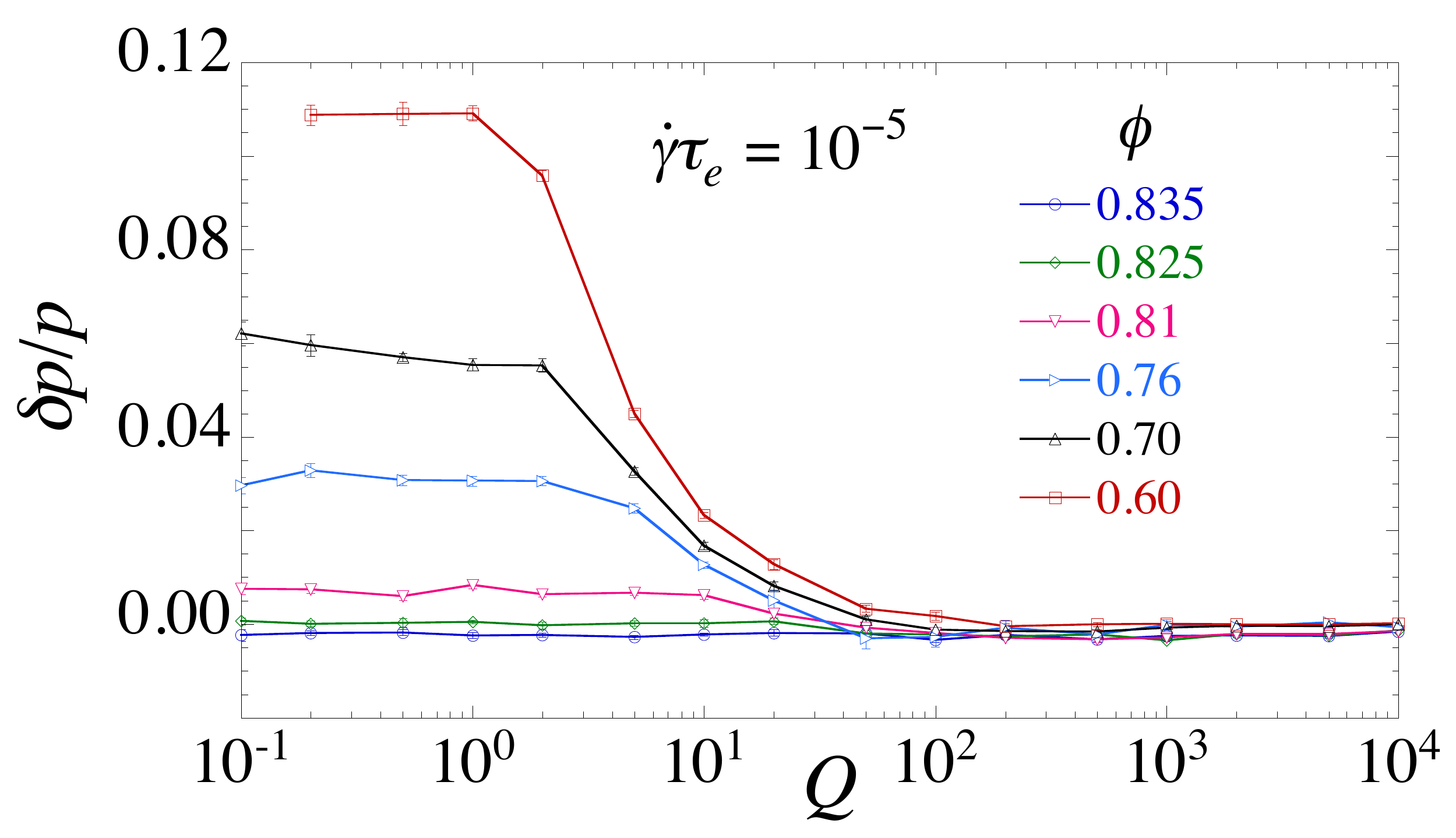}
\caption{(Color online) The relative anisotropy in pressure, $\delta p/p$, vs $Q$ for different values of packing fraction $\phi$.  The shear strain rate is $\dot\gamma\tau_e=10^{-5}$ and the system has $N=1024$ particles.  The value of $\phi$ increases as the curves go from top to bottom.
}
\label{dp-vs-Q}
\end{figure}

In Fig.~\ref{dp-vs-Q} we plot $\delta p/p$ vs $Q$ for several different packing fractions $\phi$.  Our results are for a system with $N=1024$ particles and a shear strain rate of $\dot\gamma\tau_e=10^{-5}$.
We see that for all $\phi$, $\delta p/p$ is very small at high $Q$.  However for small $Q\lesssim 10$, $\delta p/p$ can be of the order $5-12\%$ at the smaller values of $\phi$.  We find that the contribution to $\delta p$ from the dissipative part of the pressure tensor is always negligible, while the elastic part contributes roughly twice as much as the kinetic part at low $\phi$; as $\phi$ decreases, the relative contribution of the kinetic part tends to increase.

\begin{figure}[h!]
\includegraphics[width=3.2in]{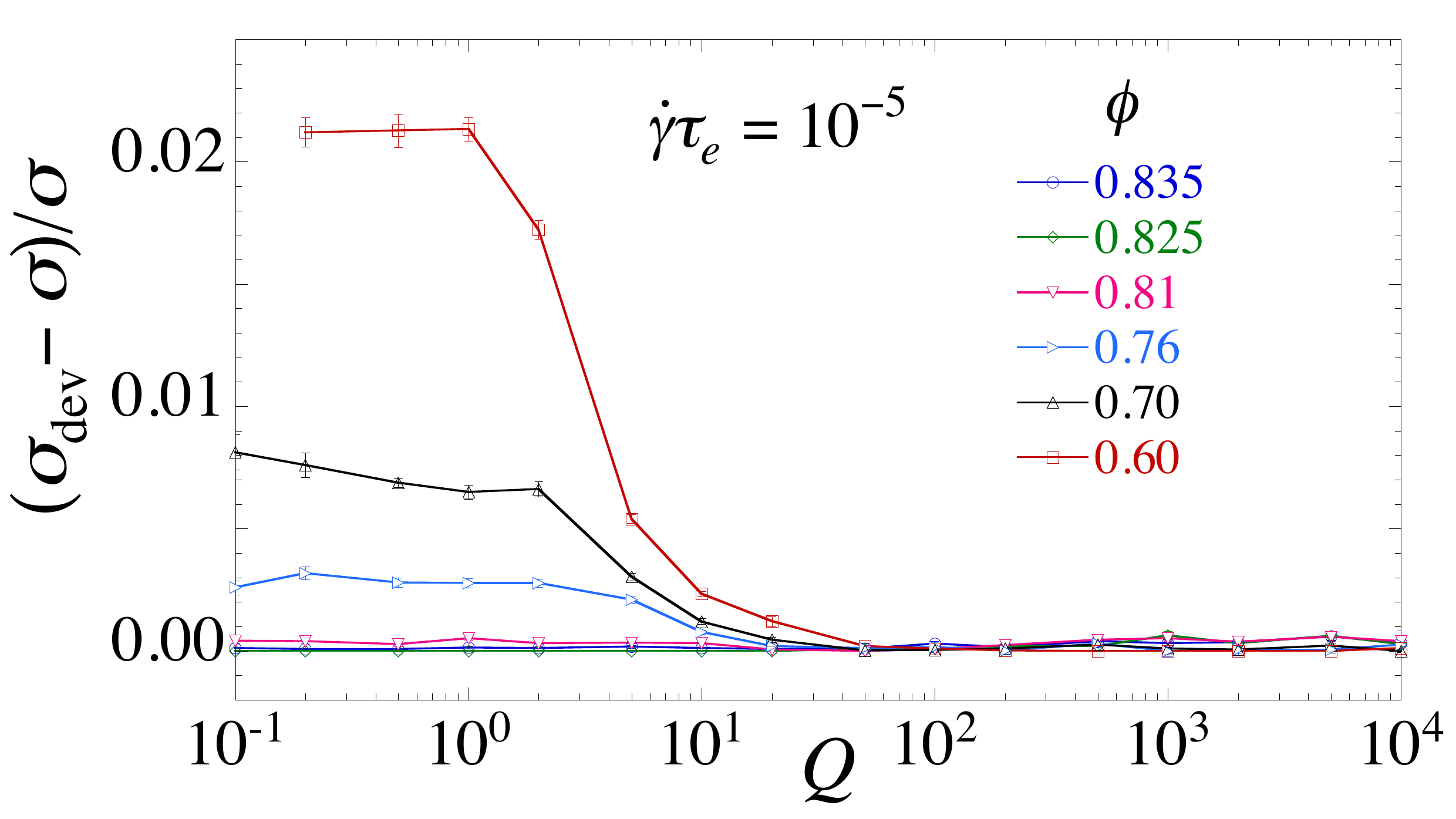}
\caption{(Color online) The relative difference between deviatoric shear stress and off-diagonal stress, $(\sigma_\mathrm{dev}-\sigma)/\sigma$, vs $Q$ for different values of packing fraction $\phi$.  The shear strain rate is $\dot\gamma\tau_e=10^{-5}$ and the system has $N=1024$ particles.  The value of $\phi$ increases as the curves go from top to bottom.
}
\label{sigdev-vs-Q}
\end{figure}

In Fig.~\ref{sigdev-vs-Q} we show the corresponding results for $(\sigma_\mathrm{dev}-\sigma)/\sigma$.  Here we see that this quantity is fairly small everywhere, reaching its largest value of  $\sim 2\%$ for the smallest $\phi=0.60$ at small $Q\lesssim 1$.  We can understand why $(\sigma_\mathrm{dev}-\sigma)/\sigma$ is small by writing,
\begin{equation}
\dfrac{\sigma_\mathrm{dev}-\sigma}{\sigma} = \sqrt{\left[ \left(\dfrac{\delta p/p}{\sigma/p}\right)^2 + 1\right]} \,-1.
\end{equation}
Comparing Fig.~\ref{dp-vs-Q} with Fig.~\ref{mu-vs-phi}, we see that where $\delta p/p$ is largest, $\sigma/p=\mu$ is also largest, with the result that the first factor under the square root is always small.  

\begin{thebibliography}{99}
\bibitem{noteMass}Note that in Ref.~\cite{VOTBagnold1} we erroneously stated that the unit of mass was taken as $m_s=1$.  In fact, the masses in that work are taken as described here, with the unit of mass $m_0=1$.  Furthermore, $\tau_e$ and $\tau_d$ in that work are also defined with respect to $m_0$, as described here, and not with respect to $m_s$ as was erroneously stated.
\bibitem{noteDt}If one is only exploring states already close to the steady state, then it is no longer necessary to employ the condition $\Delta t/\tau_e\le 0.5/Q$ at large $Q$, since near the steady state the velocity scale is set by the strain rate $\dot\gamma$ and so does not get uncontrollably large.  However this condition is useful to include when starting in states that are far from the steady state, and this is what we used in practice.
\bibitem{OHLnote2}In Ref.~\cite{OHL} the authors use an analytic expression for $\tau_\mathrm{dur}$ appropriate for the head-on collision of two particles, rather than compute $\tau_\mathrm{dur}$ directly.  We believe this may be the reason they see close, but not perfect, agreement with $\langle Z\rangle=2\tau_\mathrm{dur}\nu_\mathrm{coll}$ in their Fig.~8b.
\end{thebibliography}
\end{document}